\documentclass[aps,prd,twocolumn,showpacs,superscriptaddress,groupedaddress]{revtex4}

\usepackage{graphicx}
\usepackage{dcolumn}
\usepackage{bm}%
\usepackage{amssymb,amsmath}   
\usepackage{epsfig}
\usepackage{psfrag}
\usepackage{xspace}
\usepackage{multirow}
\usepackage{lscape}
\usepackage{comment}

\newcommand{\sigmatt}{\mbox{${\sigma}_{t\bar{t}}$}\xspace}

\newcommand{\ttbar}     {\mbox{$t\bar{t}$}\xspace}
\newcommand{\ppbar}     {\mbox{$p\bar{p}$}\xspace}
\newcommand{\eplus}	{\mbox{$e$+jets}\xspace}
\newcommand{\muplus}    {\mbox{$\mu$+jets}\xspace}
\newcommand{\lplus}     {\mbox{$\ell$+jets}\xspace}
\newcommand{\ejets}	{\mbox{$e$+jets}\xspace}
\newcommand{\mujets}    {\mbox{$\mu$+jets}\xspace}
\newcommand{\ljets}     {\mbox{$\ell$+jets}\xspace}
\newcommand{\met}       {\mbox{$\not\!\!E_T$}}
\newcommand{\pythia}    {\mbox{\textsc{pythia}}}
\newcommand{\geant}     {{\sc{geant}}}
\newcommand{\alpgen}    {\mbox{\textsc{alpgen}}}
\newcommand{\mcatnlo}    {\mbox{\textsc{mc@nlo}}}
\newcommand{\lumi}      {5.3~fb$^{-1}$}

\newcommand{\combiresult}  {7.78}

\newcommand{\combitotal}   {^{+0.77}_{-0.64}}

\newcommand{\herwig}    {\textsc{herwig}\xspace}

\newcommand{\ttb}{\mbox{$t\bar{t}$}\xspace}

\begin{document}

\hspace{5.2in} \mbox{FERMILAB-PUB-10/544-E}

\title{Measurement of the top quark pair production cross section in the lepton+jets
  channel in proton-antiproton collisions at $\sqrt{s}$=1.96 TeV}

\affiliation{Universidad de Buenos Aires, Buenos Aires, Argentina}
\affiliation{LAFEX, Centro Brasileiro de Pesquisas F{\'\i}sicas, Rio de Janeiro, Brazil}
\affiliation{Universidade do Estado do Rio de Janeiro, Rio de Janeiro, Brazil}
\affiliation{Universidade Federal do ABC, Santo Andr\'e, Brazil}
\affiliation{Instituto de F\'{\i}sica Te\'orica, Universidade Estadual Paulista, S\~ao Paulo, Brazil}
\affiliation{Simon Fraser University, Vancouver, British Columbia, and York University, Toronto, Ontario, Canada}
\affiliation{University of Science and Technology of China, Hefei, People's Republic of China}
\affiliation{Universidad de los Andes, Bogot\'{a}, Colombia}
\affiliation{Charles University, Faculty of Mathematics and Physics, Center for Particle Physics, Prague, Czech Republic}
\affiliation{Czech Technical University in Prague, Prague, Czech Republic}
\affiliation{Center for Particle Physics, Institute of Physics, Academy of Sciences of the Czech Republic, Prague, Czech Republic}
\affiliation{Universidad San Francisco de Quito, Quito, Ecuador}
\affiliation{LPC, Universit\'e Blaise Pascal, CNRS/IN2P3, Clermont, France}
\affiliation{LPSC, Universit\'e Joseph Fourier Grenoble 1, CNRS/IN2P3, Institut National Polytechnique de Grenoble, Grenoble, France}
\affiliation{CPPM, Aix-Marseille Universit\'e, CNRS/IN2P3, Marseille, France}
\affiliation{LAL, Universit\'e Paris-Sud, CNRS/IN2P3, Orsay, France}
\affiliation{LPNHE, Universit\'es Paris VI and VII, CNRS/IN2P3, Paris, France}
\affiliation{CEA, Irfu, SPP, Saclay, France}
\affiliation{IPHC, Universit\'e de Strasbourg, CNRS/IN2P3, Strasbourg, France}
\affiliation{IPNL, Universit\'e Lyon 1, CNRS/IN2P3, Villeurbanne, France and Universit\'e de Lyon, Lyon, France}
\affiliation{III. Physikalisches Institut A, RWTH Aachen University, Aachen, Germany}
\affiliation{Physikalisches Institut, Universit{\"a}t Freiburg, Freiburg, Germany}
\affiliation{II. Physikalisches Institut, Georg-August-Universit{\"a}t G\"ottingen, G\"ottingen, Germany}
\affiliation{Institut f{\"u}r Physik, Universit{\"a}t Mainz, Mainz, Germany}
\affiliation{Ludwig-Maximilians-Universit{\"a}t M{\"u}nchen, M{\"u}nchen, Germany}
\affiliation{Fachbereich Physik, Bergische Universit{\"a}t Wuppertal, Wuppertal, Germany}
\affiliation{Panjab University, Chandigarh, India}
\affiliation{Delhi University, Delhi, India}
\affiliation{Tata Institute of Fundamental Research, Mumbai, India}
\affiliation{University College Dublin, Dublin, Ireland}
\affiliation{Korea Detector Laboratory, Korea University, Seoul, Korea}
\affiliation{CINVESTAV, Mexico City, Mexico}
\affiliation{FOM-Institute NIKHEF and University of Amsterdam/NIKHEF, Amsterdam, The Netherlands}
\affiliation{Radboud University Nijmegen/NIKHEF, Nijmegen, The Netherlands}
\affiliation{Joint Institute for Nuclear Research, Dubna, Russia}
\affiliation{Institute for Theoretical and Experimental Physics, Moscow, Russia}
\affiliation{Moscow State University, Moscow, Russia}
\affiliation{Institute for High Energy Physics, Protvino, Russia}
\affiliation{Petersburg Nuclear Physics Institute, St. Petersburg, Russia}
\affiliation{Stockholm University, Stockholm and Uppsala University, Uppsala, Sweden }
\affiliation{Lancaster University, Lancaster LA1 4YB, United Kingdom}
\affiliation{Imperial College London, London SW7 2AZ, United Kingdom}
\affiliation{The University of Manchester, Manchester M13 9PL, United Kingdom}
\affiliation{University of Arizona, Tucson, Arizona 85721, USA}
\affiliation{University of California Riverside, Riverside, California 92521, USA}
\affiliation{Florida State University, Tallahassee, Florida 32306, USA}
\affiliation{Fermi National Accelerator Laboratory, Batavia, Illinois 60510, USA}
\affiliation{University of Illinois at Chicago, Chicago, Illinois 60607, USA}
\affiliation{Northern Illinois University, DeKalb, Illinois 60115, USA}
\affiliation{Northwestern University, Evanston, Illinois 60208, USA}
\affiliation{Indiana University, Bloomington, Indiana 47405, USA}
\affiliation{Purdue University Calumet, Hammond, Indiana 46323, USA}
\affiliation{University of Notre Dame, Notre Dame, Indiana 46556, USA}
\affiliation{Iowa State University, Ames, Iowa 50011, USA}
\affiliation{University of Kansas, Lawrence, Kansas 66045, USA}
\affiliation{Kansas State University, Manhattan, Kansas 66506, USA}
\affiliation{Louisiana Tech University, Ruston, Louisiana 71272, USA}
\affiliation{Boston University, Boston, Massachusetts 02215, USA}
\affiliation{Northeastern University, Boston, Massachusetts 02115, USA}
\affiliation{University of Michigan, Ann Arbor, Michigan 48109, USA}
\affiliation{Michigan State University, East Lansing, Michigan 48824, USA}
\affiliation{University of Mississippi, University, Mississippi 38677, USA}
\affiliation{University of Nebraska, Lincoln, Nebraska 68588, USA}
\affiliation{Rutgers University, Piscataway, New Jersey 08855, USA}
\affiliation{Princeton University, Princeton, New Jersey 08544, USA}
\affiliation{State University of New York, Buffalo, New York 14260, USA}
\affiliation{Columbia University, New York, New York 10027, USA}
\affiliation{University of Rochester, Rochester, New York 14627, USA}
\affiliation{State University of New York, Stony Brook, New York 11794, USA}
\affiliation{Brookhaven National Laboratory, Upton, New York 11973, USA}
\affiliation{Langston University, Langston, Oklahoma 73050, USA}
\affiliation{University of Oklahoma, Norman, Oklahoma 73019, USA}
\affiliation{Oklahoma State University, Stillwater, Oklahoma 74078, USA}
\affiliation{Brown University, Providence, Rhode Island 02912, USA}
\affiliation{University of Texas, Arlington, Texas 76019, USA}
\affiliation{Southern Methodist University, Dallas, Texas 75275, USA}
\affiliation{Rice University, Houston, Texas 77005, USA}
\affiliation{University of Virginia, Charlottesville, Virginia 22901, USA}
\affiliation{University of Washington, Seattle, Washington 98195, USA}
\author{V.M.~Abazov} \affiliation{Joint Institute for Nuclear Research, Dubna, Russia}
\author{B.~Abbott} \affiliation{University of Oklahoma, Norman, Oklahoma 73019, USA}
\author{B.S.~Acharya} \affiliation{Tata Institute of Fundamental Research, Mumbai, India}
\author{M.~Adams} \affiliation{University of Illinois at Chicago, Chicago, Illinois 60607, USA}
\author{T.~Adams} \affiliation{Florida State University, Tallahassee, Florida 32306, USA}
\author{G.D.~Alexeev} \affiliation{Joint Institute for Nuclear Research, Dubna, Russia}
\author{G.~Alkhazov} \affiliation{Petersburg Nuclear Physics Institute, St. Petersburg, Russia}
\author{A.~Alton$^{a}$} \affiliation{University of Michigan, Ann Arbor, Michigan 48109, USA}
\author{G.~Alverson} \affiliation{Northeastern University, Boston, Massachusetts 02115, USA}
\author{G.A.~Alves} \affiliation{LAFEX, Centro Brasileiro de Pesquisas F{\'\i}sicas, Rio de Janeiro, Brazil}
% ...
%
%
\collaboration{The D0 Collaboration\footnote{with visitors from
$^{a}$Augustana College, Sioux Falls, SD, USA,
$^{b}$The University of Liverpool, Liverpool, UK,
$^{c}$SLAC, Menlo Park, CA, USA,
$^{d}$ICREA/IFAE, Barcelona, Spain,
$^{e}$Centro de Investigacion en Computacion - IPN, Mexico City, Mexico,
$^{f}$ECFM, Universidad Autonoma de Sinaloa, Culiac\'an, Mexico,
and 
$^{g}$Universit{\"a}t Bern, Bern, Switzerland.%
}} \noaffiliation
\vskip 0.25cm
       
\date{December 30, 2010}

\begin{abstract} 
We present a measurement of the inclusive top quark  pair production
cross section in \ppbar ~collisions at \( \sqrt{s}=1.96 \) TeV
utilizing data corresponding to an integrated luminosity of \lumi\  collected with the D0 detector at the
Fermilab Tevatron Collider.  
We consider final states containing one high-$p_{T}$ 
 isolated electron or muon and at least two jets,  
and we perform three analyses: one exploiting specific kinematic features of
\ttbar ~events, the second using $b$-jet identification, and the third using  
both techniques to separate \ttbar\ signal from background. 
In the third case, we determine   simultaneously 
the $t\bar{t}$ cross section and the ratio of the production rates of $W$+heavy flavor jets and $W$+light flavor jets, which reduces the impact of the systematic uncertainties related to the
background estimation. Assuming a top quark mass of 172.5 GeV, we obtain
$\sigma_{t\overline{t}} = 7.78^{+0.77}_{-0.64} $ pb. This result 
agrees with predictions of the standard model.

\end{abstract}


\maketitle

\section{Introduction \label{sec:Introduction}}

The inclusive \ttb production cross section ($\sigma_{t\bar{t}}$) 
is predicted in the standard model (SM) with a 
precision of $6\%$~to~$8\%$ ~\cite{SMtheory_A,SMtheory_M,SMtheory_K,SMtheory_K2,SMtheory_C}.
Due to the large mass of the top quark, many models of physics beyond the SM predict 
observable effects in the top quark sector which can affect the top quark production rate.  
For example, the decay of a top quark into a charged Higgs boson and a
$b$~quark ($t\rightarrow H^{+}b$) would affect
the value of $\sigma_{t\bar{t}}$ extracted from 
different final states \cite{ratio,HplusD0,HplusCDF}. 
In the SM, the top quark decays with almost 100\% probability into a $W$~boson and a $b$~quark.

In this article, we present a new measurement of the inclusive top quark 
production cross section in $p\bar{p}$  collisions
at $\sqrt{s}=1.96$~TeV in the lepton+jets (\ljets) final state
where one of the 
$W$ bosons from the top quark decays hadronically into a $q\overline{q}'$ pair and the other 
leptonically into $e\nu_e$, $\mu \nu_{\mu}$, or $\tau \nu_\tau$. 
We consider both direct electron and muon decays, as well as 
secondary electrons and muons from $\tau$ decay, but not taus decaying hadronically. 
If both $W$ bosons decay leptonically, this leads to a dilepton final state 
containing a pair of electrons, a pair of muons,  or an electron and a muon, 
all of opposite electric charge. If only one of the
leptons is reconstructed, the dilepton decay chain is also included in the signal. We also include events where  both $W$ bosons decay leptonically,
and one lepton is an electron or muon and the other a hadronically decaying $\tau$ lepton.
 The 
\ttbar\ processes where both $W$ bosons decay hadronically contribute to multijet production, which is considered as a background process in this analysis.

We measure the $t\bar{t}$ production cross section using three methods: ({\it i}) a 
``kinematic'' method based on \ttbar event kinematics, ({\it ii}) a 
``counting'' method using $b$-jet identification, and ({\it iii})
a method utilizing both techniques, referred to  as the ``combined'' method. The first method does not rely on the identification of $b$~quarks while the second and third methods do. Thus they 
are sensitive to different systematic uncertainties. The  combined 
method allows the simultaneous measurement of the $t\bar{t}$ production cross section 
and of the contribution from the largest background source.   

The analysis is based on data collected with the D0 detector~\cite{d0det} in Run II of the Fermilab Tevatron Collider   with an
integrated luminosity of  
 $5.3\pm 0.3$~${\rm fb^{-1}}$.  The results of this analysis supersede our previous measurement~\cite{PRLxs}, which was done with a fifth of the dataset considered here. 
A result from the CDF Collaboration is available in Ref.~\cite{cdf_xsec}.
Recently, the ATLAS and CMS Collaborations reported first measurements of the \ttbar cross section in
  $pp$  collisions at $\sqrt{s}=7.0$~TeV~\cite{atlas,cms}.  

In 2006,  
the D0 detector was substantially upgraded: 
a new calorimeter trigger was installed~\cite{L1cal}
and a new inner layer was added to the
silicon microstrip tracker~\cite{layer0}.   
We split the data into two samples: Run IIa before this upgrade (on which our previous $t\bar{t}$ cross section measurement was performed) and Run IIb 
after it. The corresponding integrated luminosities are 1~fb$^{-1}$ and 
4.3~fb$^{-1}$, respectively.

\section{D0 Detector}
The D0 detector contains a tracking system, a calorimeter, and a muon
spectrometer~\cite{d0det}. The tracking system consists of a silicon microstrip
tracker (SMT) and a central fiber tracker (CFT), both located inside a 1.9~T
superconducting solenoid.  
The design provides efficient charged-particle tracking in the detector 
pseudorapidity region $|\eta_{det}| < 3$~\cite{eta}. The
SMT provides the capability to reconstruct the
$p\bar{p}$ interaction vertex (PV) with a precision of about 40 $\mu m$ in the
plane transverse to the beam direction,  
and to determine the impact
parameter of any track relative to the PV~\cite{ip} with a precision between 
20 and 50 $\mu \rm{m}$, depending on the number of hits in the SMT,  
which is key to lifetime-based $b$-jet tagging.
The calorimeter
has a central section covering  $|\eta_{det}|<1.1$, and two end
calorimeters (EC) extending the coverage to  $|\eta_{det}|\approx 4.2$.  The
muon system 
surrounds the calorimeter and consists of three layers of tracking detectors 
and  scintillators covering  $|\eta_{det}|<2$~\cite{muon_detector}. 
A 1.8~T toroidal 
iron magnet is located outside the innermost layer of the muon detector.
The luminosity is calculated from the rate of 
{\mbox{$p\bar p$}}\ inelastic collisions measured with plastic
scintillator arrays, which are located in front of the EC
cryostats.

The D0 trigger is based on a three-level pipeline system. The first level consists 
of hardware and firmware components. The microprocessor-based second level combines information from the 
different detector components to construct simple physics objects, whereas the software-based third level 
uses the full event information obtained with a simplified reconstruction \cite{matrixmethod}.

\section{Event selection}
\label{sec:eventsel}
Events in the l+jets channel are triggered by requiring
either an electron or a lower-$p_T$ electron accompanied by a jet
for the \eplus\ channel, a muon and a jet for the \muplus\  final state in
Run~IIa, and a muon for the \muplus final state in Run IIb.
These samples are
enriched in $t\bar{t}$ events by  
requiring more than one jet of cone radius ${\cal R}=0.5$ \cite{DeltaR}
reconstructed with the ``Run II cone'' algorithm~\cite{RunIIcone}, 
with transverse momentum $p_T>20$~GeV 
and pseudorapidity $|\eta_{det}|<2.5$. Furthermore, we require 
one isolated electron with $p_T>20$~GeV and $|\eta_{det}|<1.1$, or one
isolated muon 
with $p_T>20$~GeV and $|\eta_{det}|<2.0$, and missing transverse energy 
$\met>20 (25)$~GeV in the \ejets\ (\mujets)
channel. The PV must be within $60$~cm of the detector
center in the longitudinal coordinate so that it is within the SMT fiducial region. 
In addition, the jet with highest $p_T$ must have   
$p_T > 40$~GeV. The high instantaneous luminosity achieved by the Tevatron 
leads to a significant contribution from additional \ppbar\ collisions 
within the same bunch crossing as the hard interaction. To reject jets from these additional collisions, we require all jets in Run~IIb to contain at least three 
tracks within each jet cone that originate from the PV. Events containing two isolated leptons (either $e$ or $\mu$) 
with $p_T>15$~GeV are rejected.

The $b$-jets are identified using a neural network formed by combining variables characterizing the properties of secondary vertices and of
     tracks with large impact parameters relative to the PV~\cite{btagging}. 
Details of lepton identification, jet identification and missing transverse energy calculation are described in Ref.~\cite{matrixmethod}. 

We split the selected \ljets\ sample into subsamples according 
to lepton flavor ($e$ or $\mu$) and jet multiplicity, and between Run IIa and Run IIb.
For the measurements with $b$-tagging, we split the data into additional 
subsamples according to the
number of tagged $b$-jet candidates (0, 1 or $> 1$).

\section{Sample composition}
\label{sec:samplecompo}
Top quark pair production and decay 
is simulated with the {\alpgen} Monte Carlo (MC) program~\cite{Alpgen} 
assuming a top quark mass of $m_t=172.5$~GeV (used for all tables and figures in this paper unless stated otherwise). 
The fragmentation of partons and the hadronization process
are simulated using  {\pythia}~\cite{Pythia}. A matching scheme is
applied to avoid double-counting of partonic event configurations~\cite{matching}.
The generated events are processed through a
\geant-based ~\cite{geant} simulation of the D0 detector and the same
reconstruction programs used for the data. Effects from additional $p\bar{p}$ interactions are simulated by overlaying data from random
 $p\bar{p}$ crossings over the MC events. 

The background can be split into two components:  
``instrumental background,'' where the decay products of a
final state parton are reconstructed as an isolated lepton,
and ``physics background'' that originates from processes
with a final state similar to that of $t\bar{t}$ signal.
In the \eplus\ channel, instrumental background arises from multijet (MJ) production 
when a jet with high electromagnetic content mimics an electron;  
in the \muplus channel, it occurs when a muon contained within a jet
originates from the decay of a heavy-flavor quark ($b$ or $c$ quark), but appears isolated. 

The dominant physics background is from
$W$+jets production. Other physics backgrounds are  
single top quark, diboson, and $Z$+jets production with $Z\rightarrow \tau\tau$, and   
$Z\rightarrow ee$ ($Z\rightarrow \mu\mu$)  in the \eplus\ (\muplus)
channel. The contributions from these background sources are estimated using MC simulations and 
normalized to next-to-leading order (NLO) predictions.
Diboson events ($WW$, $WZ$ and $ZZ$) are generated with {\pythia}, single top quark
production with the {\textsc{comphep}} generator~\cite{CompHep}, and
$Z$+jets events, with $Z\rightarrow ee$, $\mu\mu$, and $\tau\tau$, are
simulated using {\alpgen}. For the $Z$+jets background, the $p_T$ distribution of 
the $Z$ boson is 
corrected to match the distribution observed in data, taking into account
a dependence on jet multiplicity. 
All simulated samples are generated using the CTEQ6L1 parton distribution functions~(PDFs)
\cite{PDF}. The main background contribution, which is $W$+jets events, is discussed further below. 

The MJ background is estimated from data 
using the ``matrix method''~\cite{matrixmethod}: Two samples of \lplus\ events are designed categorized by  
the stringency of the lepton selection criteria: the ``tight'' sample used for the signal extraction is a subset of a  ``loose'' set which is dominated by background. 
The number of MJ events is extracted using event counts in 
these two samples and the corresponding  isolated  lepton reconstruction and identification efficiencies ($\epsilon_{s}$) and
the probability of misidentifying a jet as a lepton ($\epsilon_{b}$), determined for Run IIa
and Run IIb data separately. 
The efficiency $\epsilon_{b}$ is measured in a sample of events 
that pass the same selection as the signal sample, but has 
low \met. This sample is dominated by MJ 
events, and the remaining contributions from  isolated leptons are subtracted. 
The efficiency $\epsilon_{s}$ is extracted from $W$+jets and $t\bar{t}$ MC events
calibrated to reproduce lepton reconstruction and identification efficiencies 
in data.   
Neither $\epsilon_{b}$ nor $\epsilon_{s}$ shows any statistically 
significant dependence 
on the jet multiplicity, and both are obtained from a sample with 
at least two jets. 
Table~\ref{tab:epsMM} shows the measured values of $\epsilon_{s}$ and $\epsilon_{b}$ 
for Run IIa and Run IIb, and Table \ref{tab:numMM} provides the numbers 
of selected ``loose'' and ``tight'' events in each jet multiplicity bin. 
  The kinematic distributions for the MJ background are obtained from 
the \lplus\ data sample of loose leptons that do not 
fulfill the tight isolation criteria.  

\begin{table}[ht]
\begin{center}
\caption{Efficiencies for isolated leptons and misidentified jets  to pass the tight 
selection criteria. The uncertainties include both statistical and systematic contributions. 
\label{tab:epsMM}}
\begin{tabular}{lcc}
\hline\hline
 & \eplus & \muplus \\ \hline
& \multicolumn{2}{c}{Run IIa}   \\ \hline
$\varepsilon_{s}$ & $0.831 \pm 0.011$ & $0.881 \pm 0.039$\\
$\varepsilon_{b}$ &  $ 0.109\pm 0.008$ &  $0.172\pm 0.048$ \\ \hline 
& \multicolumn{2}{c}{Run IIb}   \\ \hline
$\varepsilon_{s}$ & $0.813 \pm 0.045$ & $0.896 \pm 0.021$\\
$\varepsilon_{b}$ &  $ 0.124\pm 0.015$ &  $0.219\pm 0.043$ \\ \hline\hline
\end{tabular}
\end{center}
\end{table}

\begin{table*}[ht]
\begin{center}
\begin{minipage}{5.5 in}
\caption{Numbers of selected ``loose'' ($N_L$) and ``tight'' ($N_T$) events used as 
input for the MJ background estimate as a function of jet multiplicity for samples before and after applying the $b$-tagging criteria.  
\label{tab:numMM}}
\begin{ruledtabular}
\begin{tabular}{lccc|ccc}
 & \multicolumn{3}{c}\eplus & \multicolumn{3}{c}\muplus \\ \hline
 \multicolumn{7}{c}{Run IIa}   \\ \hline
&  2 jets & 3 jets & $>$3 jets & 2 jets & 3 jets & $>$3 jets \\  \hline
$N_{L}$ & 16634 & 4452 & 1109 & 7198 & 1751 & 516 \\  
$N_{T}$ &  7649 & 1681 & 448  &  5905 & 1360 & 390 \\ \hline
$N_{L}$ 1 $b$-tag  & 996 & 450 & 196 & 413 & 187 & 129\\ 
$N_{T}$ 1 $b$-tag  & 453 & 198 & 112 &  317 & 140 & 109 \\\hline
$N_{L}$ $>$1 $b$-tag & 73 & 78 & 45 & 33 & 45 & 38 \\ 
$N_{T}$ $>$1 $b$-tag  &  48 & 45 & 33 & 28 & 38 & 35    \\\hline
 \multicolumn{7}{c}{Run IIb}   \\ \hline
&  2 jets & 3 jets & $>$3 jets & 2 jets & 3 jets & $>$3 jets \\  \hline
$N_{L}$       & 37472 & 8153 & 1914 & 17581 & 3457 & 925  \\  
$N_{T}$       & 20423 & 4118 & 1012 &  15290 & 2904 & 783 \\\hline
$N_{L}$ 1 $b$-tag  &  2917 & 1130 & 465  & 1364 & 506 & 278 \\  
$N_{T}$ 1 $b$-tag  &  1590 & 648  & 289  & 1139 & 426 & 236 \\\hline
$N_{L}$ $>$1 $b$-tag  &  251  & 218  & 164  & 125 & 126 & 127  \\  
$N_{T}$ $>$1 $b$-tag  &  184  & 154  & 127  & 109 & 114 & 119 \\
\end{tabular}
\end{ruledtabular}
\end{minipage}
\end{center}
\end{table*}

In $W$+jets production, the $W$ boson is produced through
the electroweak interaction, and additional partons are generated by
QCD radiation. Several MC generators are capable of performing matrix element 
calculations for $W$ boson production including one or more partons in
the final state, however these are performed only at tree
level. Therefore, the overall 
normalization suffers from large theoretical
uncertainties. For this reason, 
only the differential distributions are taken from the simulation while  
the overall  normalization of the $W$+jets background is obtained
from data by subtracting the physics and instrumental backgrounds
and the $t\bar{t}$ signal.
This is done as a function of jet multiplicity for each of the
analysis channels. 
The $W$+jets contribution is divided into three exclusive  
categories according to parton flavor: 
(i) ``$W+{\rm hf}$"  is the sum of all $W$ events with a $b\bar{b}$ of $c\bar{c}$ quark  pair and any number of additional jets; 
(ii) ``$W+c$" has events with a $W$ boson  produced with a single  charm quark and any number of additional jets; 
and (iii) ``$W+{\rm lf}$" has  $W$ bosons that are produced with light flavor jets.   
These three processes are generated by the LO QCD generator {\alpgen}. The
relative contributions from the three classes of events are determined using NLO
QCD calculations based on the MCFM MC generator~\cite{mcfm}. We correct
the $W+{\rm hf}$ ($W+c$) rate obtained from {\alpgen} by a K-factor of
 $1.47 \pm 0.22$
($1.27 \pm 0.15$) relative to the $W+{\rm lf}$ rate.

We verify the factor ($f_H$) which needs to be applied to the
LO $W+{\rm hf}$ rate in control samples which use the same
selection  criteria as for the signal sample, but require exactly one or exactly
two jets. To extract $f_H$, we split the events into    samples without a 
$b$-tagged jet and with at least one $b$-tagged jet, and adjust $f_H$  iteratively until
the prediction matches the data.
The resulting $f_H$ value is consistent  with the above NLO K-factor from MCFM. 
In the combined method, 
we measure $f_H$ 
  (assuming the same factor for the $b\bar{b}$ or $c\bar{c}$ components
   of $W+{\rm hf}$)
simultaneously with the $t\bar{t}$ cross section. This reduces
the uncertainties on the measured \sigmatt and provides
a measurement of this factor including the 
systematic uncertainties.

\section{\boldmath Efficiencies and Yields of ${\bm t\bar{t}}$ events}
\label{sec:efficiencyyield}

Selection efficiencies and $b$-tagging
probabilities for each of the \ttb \lplus\ channels are summarized in
Tables~\ref{tab:ljets_eff} and~\ref{tab:ljets_btag}, respectively. To calculate 
these efficiencies, we separate the 
\lplus\ $t\bar{t}$ MC events where only one $W$ boson decays 
to $e$ or $\mu$ from the dilepton $t\bar{t}$ events where both $W$ 
bosons decay leptonically, but only one lepton is reconstructed. 

We apply the same $b$-tagging algorithm to data and to simulated events, 
but correct the simulation as a function of jet  
flavor, $p_T$, and $\eta$ to achieve the same performance for $b$-tagging 
as found in data. These correction factors~\cite{btagging} are determined 
from data control samples, and are used to predict the yield   
of signal and background events with 0, 1, and $> 1$ $b$-tagged jets. 
We also correct 
lepton and jet identification and reconstruction efficiencies in simulation to 
match those measured in data.

\begin{table*}[ht]
\newcommand\T{\rule{0pt}{2.6ex}}
\newcommand\B{\rule[-1.2ex]{0pt}{0pt}}
\begin{center}
\begin{minipage}{5.5 in}
\caption{Selection efficiencies for \ttbar\ \lplus\ and dilepton 
contributions to the \lplus\ channels. The uncertainties on  the efficiencies from limited 
MC statistics are of the order of (1--2)\%.   
\label{tab:ljets_eff} }
\begin{ruledtabular}
\begin{tabular}{lcccccc}
   &   \multicolumn{3}{c} {$e$+jets}   &  \multicolumn{3}{c} {$\mu$+jets} \\ [1pt]                                                                                                                                                                                                    \hline	   			    
   & 2 jets  & 3 jets &  $>3$ jets & 2 jets  & 3 jets &  $>3$ jets \\ [1pt] \hline \\[-7pt] 	   			    
$\ttbar \to \ell$ + jets     & $0.043$ & $0.103$ & $0.097$ & $0.026$ & $0.069$ & $0.070$ \\[2pt]
$\ttbar \to \ell\ell$ + jets & $0.108$ & $0.040$ & $0.009$ & $0.067$ & $0.027$ & $0.006$ 
\end{tabular}
\end{ruledtabular}
\end{minipage}
\end{center}
\end{table*}

\begin{table*}[ht]
\begin{center}
\begin{minipage}{5.5 in}
\caption{$b$-tagging probabilities for \ttbar\ \lplus\ and dilepton 
contributions to the \lplus\ channels. The uncertainties on the $b$-tag probabilities from limited 
MC statistics are of the order of (1--2)\%.   
\label{tab:ljets_btag} }
\begin{ruledtabular}
\begin{tabular}{lcccccc}
   &   \multicolumn{3}{c} {$e$+jets}   &  \multicolumn{3}{c} {$\mu$+jets} \\  [1pt]                                                                                                                                                                                                   
\hline	   			    
   & 2 jets  & 3 jets &  $>3$ jets & 2 jets  & 3 jets &  $>3$ jets \\[1pt] \hline \\[-7pt] 
\multicolumn{7}{c} {\ttbar single tagging probabilities} \\ \hline\\[-7pt]
$\ttbar \to \ell$ + jets         & $0.431$ & $0.470$ & $0.458$ & $0.417$ & $0.464$ & $0.458$   \\
$\ttbar \to \ell\ell$ + jets     & $0.470$ & $0.459$ & $0.460$ & $0.461$ & $0.456$ & $0.438$  \\[1pt]\hline \\[-7pt]
\multicolumn{7}{c} {\ttbar double tagging probabilities } \\ [1pt] \hline \\[-7pt] 
$\ttbar \to \ell$ + jets         & $0.068$ & $0.173$ & $0.259$ & $0.066$ & $0.176$ & $0.258$    \\[1pt]
$\ttbar \to \ell\ell$ + jets     & $0.205$ & $0.241$ & $0.249$ & $0.206$ & $0.246$ & $0.271$  
\end{tabular}
\end{ruledtabular}
\end{minipage}
\end{center}
\end{table*}

Table~\ref{tab:yields_ljets} summarizes
the predicted background and the observed numbers of events in \eplus\ and
\muplus\ data with $0$, $1$, and $> 1$ tags,
together with the  prediction for the number of \ttbar\ event candidates obtained assuming the production cross section measurement from  the combined method.

\begin{table}[ht]
\caption{Yields for \eplus\ and \muplus\ with $0$, $1$, and $>1$ 
$b$-tagged jets. The number of \ttb\ events is calculated using the cross section $\sigma_{t\bar{t}}=7.78$~pb measured by the combined method. 
Uncertainties include statistical and systematic contributions. Due to the correlations of the systematic uncertainties between the samples, the uncertainty on the total predicted yield is not the sum of the uncertainties of the individual contributions. \label{tab:yields_ljets} }
\begin{center}
\begin{tabular}{l  l  r@{ $\pm$ }l  r@{ $\pm$ }l  r@{ $\pm$ }l }
\hline
      \hline
 Channel   & Sample & \multicolumn{2}{c}{0 $b$-tags} &
 \multicolumn{2}{c}{1 $b$-tag } & \multicolumn{2}{c}{$>$ 1 $b$-tags} \\
\hline 

e+2\,jets & $W$+jets   & 21019 & 517 & 1360 & 90 & 101 & 13 \\
          & Multijet   &  2531 & 301 &  197 & 25 &   6 &  1 \\
          & $Z$+jets   &  1169 & 159 &   68 & 15 &   5 &  2 \\
          & Other      &   858 &  85 &  148 & 19 &  21 &  3 \\
          & $t\bar{t}$ &   245 &  22 &  265 & 22 &  79 &  9 \\
          & Total      & 25821 & 458 & 2038 & 97 & 213 & 18 \\
          & Observed   & \multicolumn{2}{c}{25797} & \multicolumn{2}{c}{2043} &  \multicolumn{2}{c}{232} \\ 
\hline

e+3\,jets & $W$+jets   & 3358 & 151 & 316 & 26 &  29 &  4 \\
          & Multijet   &  675 &  70 &  75 &  8 &   7 &  1 \\
          & $Z$+jets   &  271 &  40 &  26 &  6 &   2 &  1 \\
          & Other      &  172 &  18 &  41 &  6 &   9 &  1 \\
          & $t\bar{t}$ &  289 &  27 & 381 & 30 & 147 & 14 \\
          & Total      & 4765 & 124 & 839 & 37 & 194 & 16 \\
          & Observed   & \multicolumn{2}{c}{4754} & \multicolumn{2}{c}{846} &  \multicolumn{2}{c}{199} \\ 
\hline

e+$>3$\,jets & $W$+jets   & 440 & 73 &  55 & 10 &   6 &  1 \\
          & Multijet   & 141 & 15 &  23 &  3 &   2 &  0 \\
          & $Z$+jets   &  43 &  7 &   6 &  2 &   1 &  0 \\
          & Other      &  30 &  4 &   8 &  1 &   2 &  0 \\
          & $t\bar{t}$ & 202 & 24 & 322 & 31 & 180 & 19 \\
          & Total      & 857 & 51 & 413 & 25 & 190 & 18 \\
          & Observed   & \multicolumn{2}{c}{899} & \multicolumn{2}{c}{401} &  \multicolumn{2}{c}{160} \\ 
\hline

$\mu$+2\,jets & $W$+jets & 17386 & 321 & 1081 & 69 &  81 & 10 \\
          & Multijet     &   208 & 117 &   38 & 24 &   1 &  1 \\
          & $Z$+jets     &  1142 & 155 &   68 & 15 &   5 &  2 \\
          & Other        &   682 &  67 &  118 & 15 &  17 &  2 \\
          & $t\bar{t}$   &   155 &  14 &  163 & 14 &  50 &  6 \\
          & Total        & 19573 & 235 & 1468 & 77 & 154 & 14 \\
          & Observed     & \multicolumn{2}{c}{19602} & \multicolumn{2}{c}{1456} &  \multicolumn{2}{c}{137} \\ 
\hline

$\mu$+3\,jets & $W$+jets & 2895 & 100 & 261 & 20 &  24 &  3 \\
          & Multijet     &   87 &  29 &  14 &  5 &   0 &  0 \\
          & $Z$+jets     &  222 &  31 &  19 &  5 &   2 &  1 \\
          & Other        &  138 &  14 &  32 &  4 &   7 &  1 \\
          & $t\bar{t}$   &  198 &  18 & 262 & 21 & 103 & 10 \\
          & Total        & 3540 &  77 & 589 & 28 & 136 & 12 \\
          & Observed     & \multicolumn{2}{c}{3546} & \multicolumn{2}{c}{566} &  \multicolumn{2}{c}{152} \\ 
\hline

$\mu+>3$\,jets & $W$+jets & 481 & 53 &  63 &  8 &   7 &  2 \\
          & Multijet     &  27 &  9 &   6 &  2 &   0 &  0 \\
          & $Z$+jets     &  29 &  5 &   4 &  1 &   1 &  1 \\
          & Other        &  23 &  3 &   7 &  1 &   2 &  0 \\
          & $t\bar{t}$   & 151 & 17 & 240 & 22 & 135 & 14 \\
          & Total        & 711 & 39 & 318 & 17 & 145 & 14 \\
          & Observed     & \multicolumn{2}{c}{674} & \multicolumn{2}{c}{345} &  \multicolumn{2}{c}{154} \\ 
\hline

      \hline
\end{tabular}
\end{center}
\end{table}

\section{Kinematic Method} \label{sec:topoxsec}
In this and the following sections we present the methods used to measure
the $t\bar{t}$ cross section. The results of the three methods are
presented in Sec.~\ref{sec:results}, after a discussion of 
the sources of systematic uncertainties in Sec.~\ref{sec:systs}.

\subsection{Discrimination}
In the kinematic analysis, we use final states with 2, 3 or $> 3$
jets, thereby defining twelve disjoint data sets.
To distinguish \ttbar\ signal from background, we 
construct a discriminant  
that exploits differences between kinematic 
properties of \ttbar\ \lplus\ signal and the dominant $W+$jets 
background using the multivariate analysis toolkit {\sc tmva}~\cite{TMVA}. 
The multivariant discriminant function is calculated by a random 
forest (RF) of decision trees. We 
use 200 trees for the RF, with the boosting type~\cite{boosting} 
set to ``bagging,'' and 
separation mode set to the ``Gini index'' without pruning~\cite{gini}. 

We split both the \ttbar\ and the $W$+jets MC events 
into two equal samples, and use one for
training and testing of the RF discriminant and the other to 
create discriminant distributions (templates) for fits to data. 
For all other sources of events,  
we use the trained RF discriminant to obtain the templates.  

We choose input variables that separate  signal and background 
and are well 
described by the MC simulation. To reduce the sensitivity of variables 
that are based on  the jets in the events to  
the modeling of soft gluon radiation and to the underlying event, 
we include only the five highest-$p_T$ (leading) jets in these definitions. The variables chosen as inputs 
to build the RF discriminant are: 

\begin{description}
\item[{\it Aplanarity}\,:]
The normalized quadratic momentum tensor ${\mathcal M}$  is defined as
\begin{equation*}
{{\mathcal M}_{ij} = \frac{\Sigma_o p^o_ip^o_j}{\Sigma_o|\vec{p}^o|^2} }  \,\, ,
\end{equation*}
where $\vec{p}^{\,o}$ is the momentum vector of a reconstructed object $o$,
and $i$ and $j$ are the three Cartesian coordinates. The sum over objects
includes up to the first five jets, ordered by $p_T$, and the selected charged lepton.
The diagonalization of
${\mathcal M}$ yields three eigenvalues
$\lambda_1 \ge \lambda_2 \ge \lambda_3$, with  $\lambda_1 + \lambda_2 +
\lambda_3=1$, that characterize the topological distribution of objects in an event.  \\ 
The {\it aplanarity} is defined as ${\mathcal A}=\frac{3}{2}\lambda_3$ and
reflects the degree of isotropy of an event, with its range restricted to 
$0 \le {\mathcal A} \le 0.5$. Large values 
correspond to spherically distributed events and small values to more planar events.
While $t\bar{t}$ final states are more spherical, as is typical for
decays of massive objects, $W+$jets and MJ events tend to be more planar.
\item[{\it Sphericity}\,:]
The {\it sphericity} is defined as ${\mathcal S}=\frac{3}{2}(\lambda_2 + \lambda_3$), and 
$t\bar{t}$ events tend to have higher values of 
${\mathcal S}$ than background events. Values of $S$ range from zero to one. 
\item[$\it{ H_T^\ell}$:] The scalar sum of the
transverse momenta of up to five leading jets ($H_T$) and the transverse momentum of the lepton.
\item[$\it{ H_T^3}$:]
The  $p_T$ of the third jet or the scalar sum of the $P_T$ of the jets with the third and fourth, or
 third to fifth largest  $p_T$ in the event, for events
 with three, four, or more jets, respectively.
As these jets 
correspond largely to gluon radiation for the 
$W$+jets background events but mainly to $W$ decays  
in the $t\bar{t}$ production, on average $\it{ H_T^3}$ has higher values for the latter process.

\item[$\it{M_T^{\rm jet}}$:] The transverse mass of the dijet system 
for $\ell+2$ jets events. Since  $\it{ H_T^3}$ is not defined in  $\ell+2$ jets events, we use $\it{M_T^{\rm jet}}$ in this channel instead.
\item[$\it{ M_{\rm event}}$:]
The invariant mass of the system consisting of the 
lepton, the neutrino and up to five leading jets.  
The energy of the neutrino is determined by
constraining the invariant mass of the lepton and vector \met\ (as the neutrino) 
to the mass of the $W$ boson. Of the two possible solutions for the
longitudinal momentum of the neutrino, we use the one with the  smaller
absolute value. On average, $\it{ M_{\rm event}}$ is larger for $t\bar{t}$ events
than for background. 
\item[$\it{ M_T^{j_2 \nu \ell}}$:]
Transverse mass of the system consisting of the second leading jet, the
lepton and the neutrino, where the energy of the neutrino is determined 
the same way as in the case of $\it{ M_{\rm event}}$. 
\end{description}

\begin{figure*}[ht]
\begin{center}
\begin{minipage}{17.0 cm}
\includegraphics[width=8.0cm]{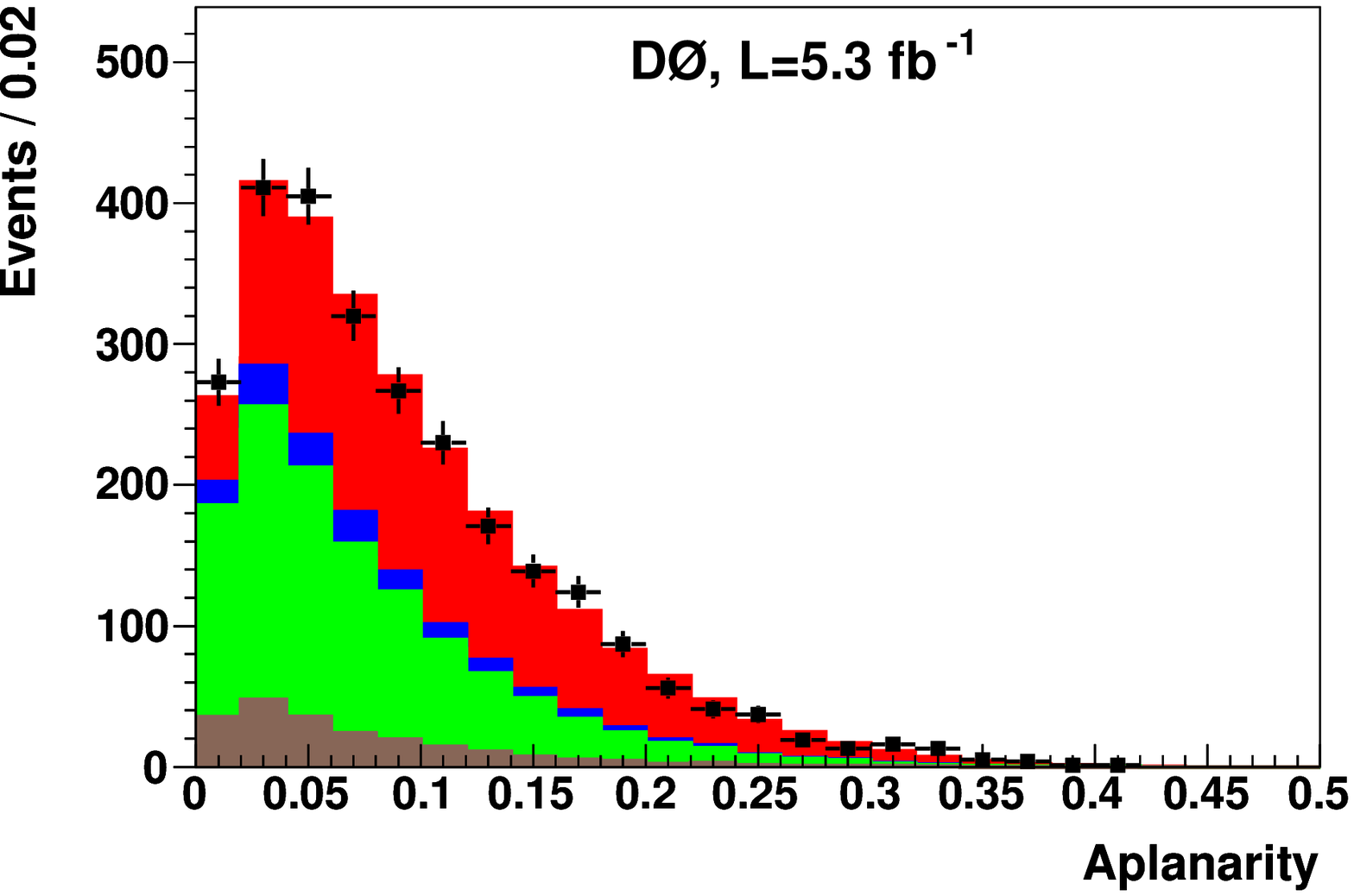}
\includegraphics[width=8.0cm]{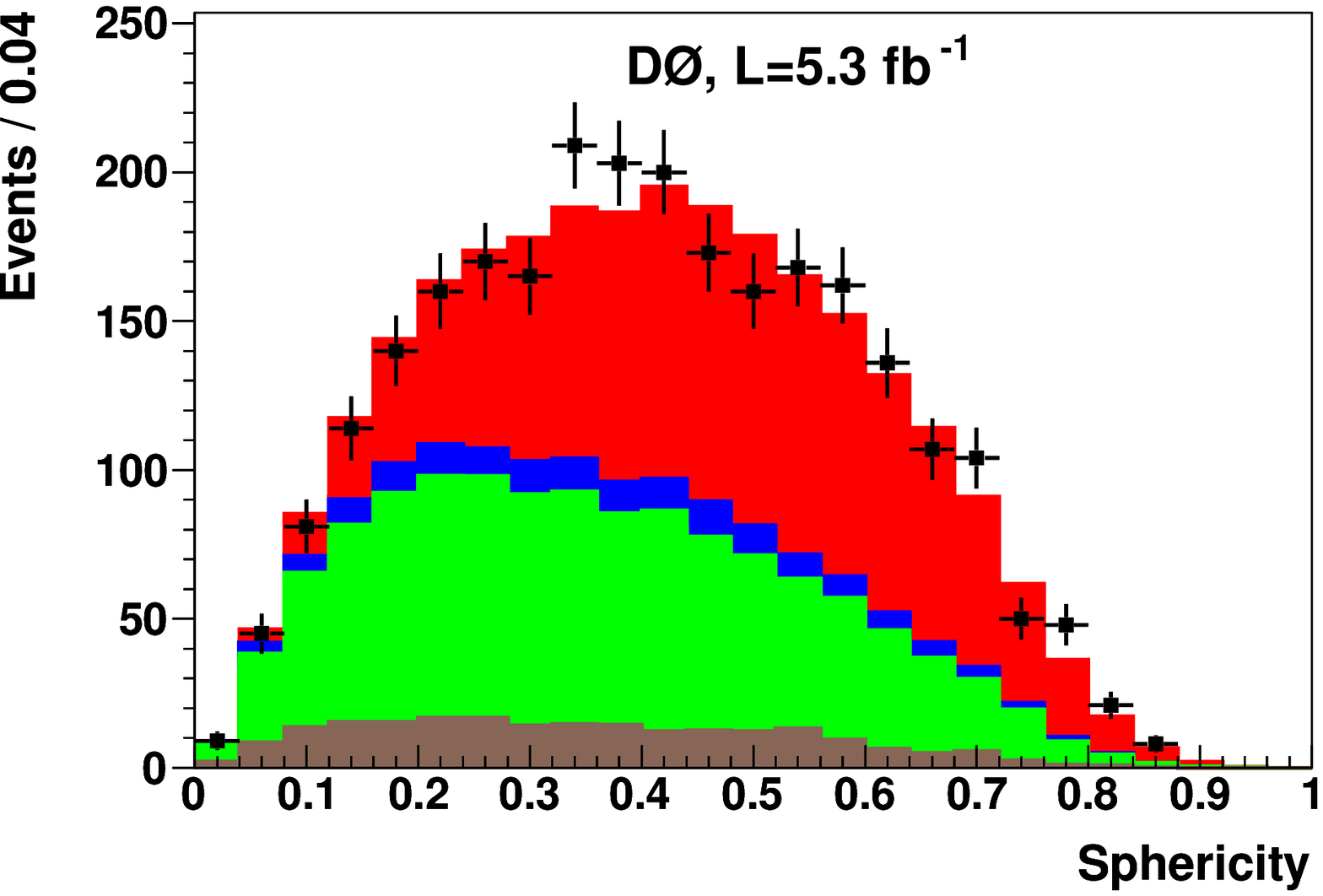}

\includegraphics[width=8.0cm]{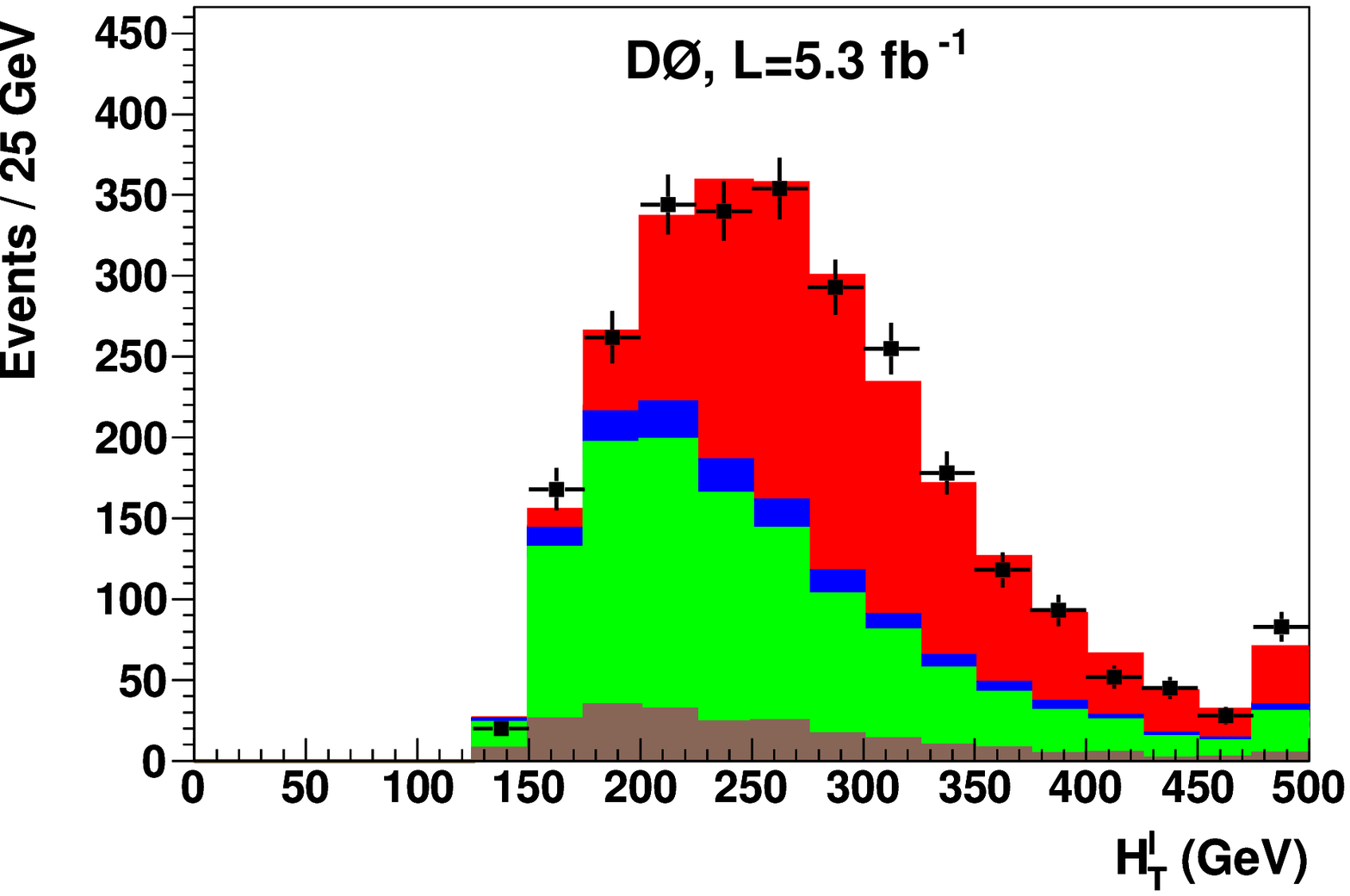}
\includegraphics[width=8.0cm]{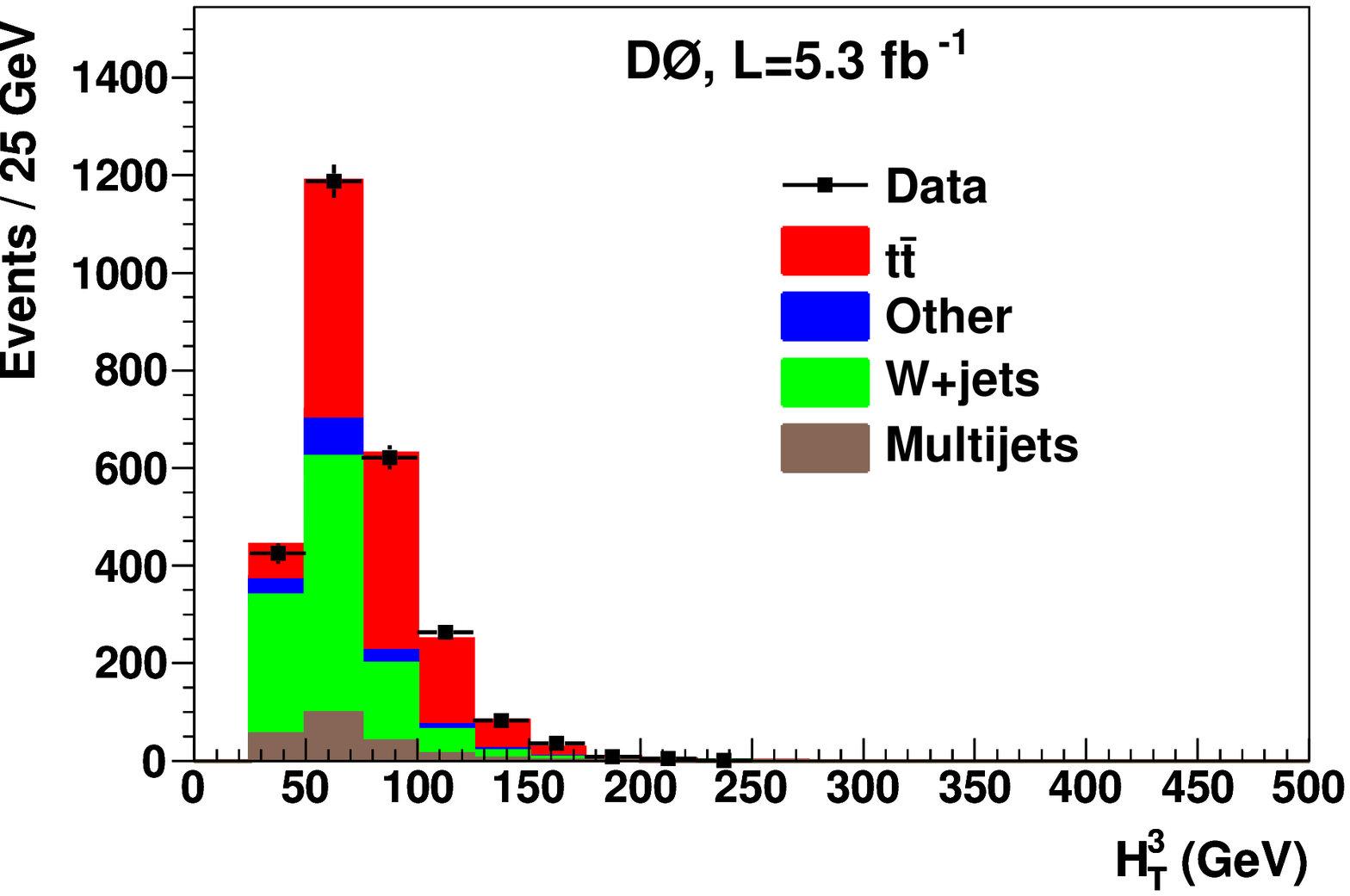}

\includegraphics[width=8.0cm]{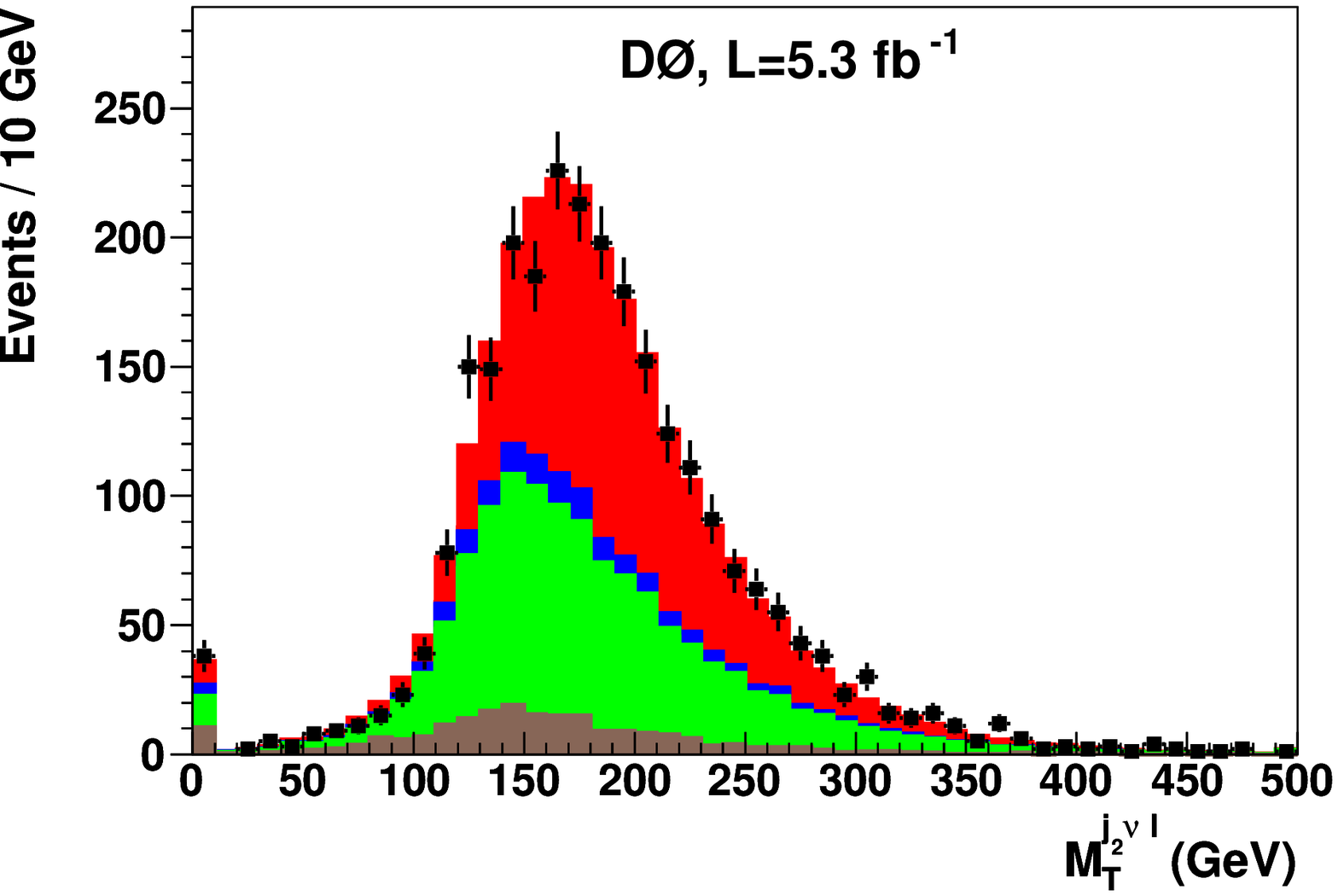}
\includegraphics[width=8.0cm]{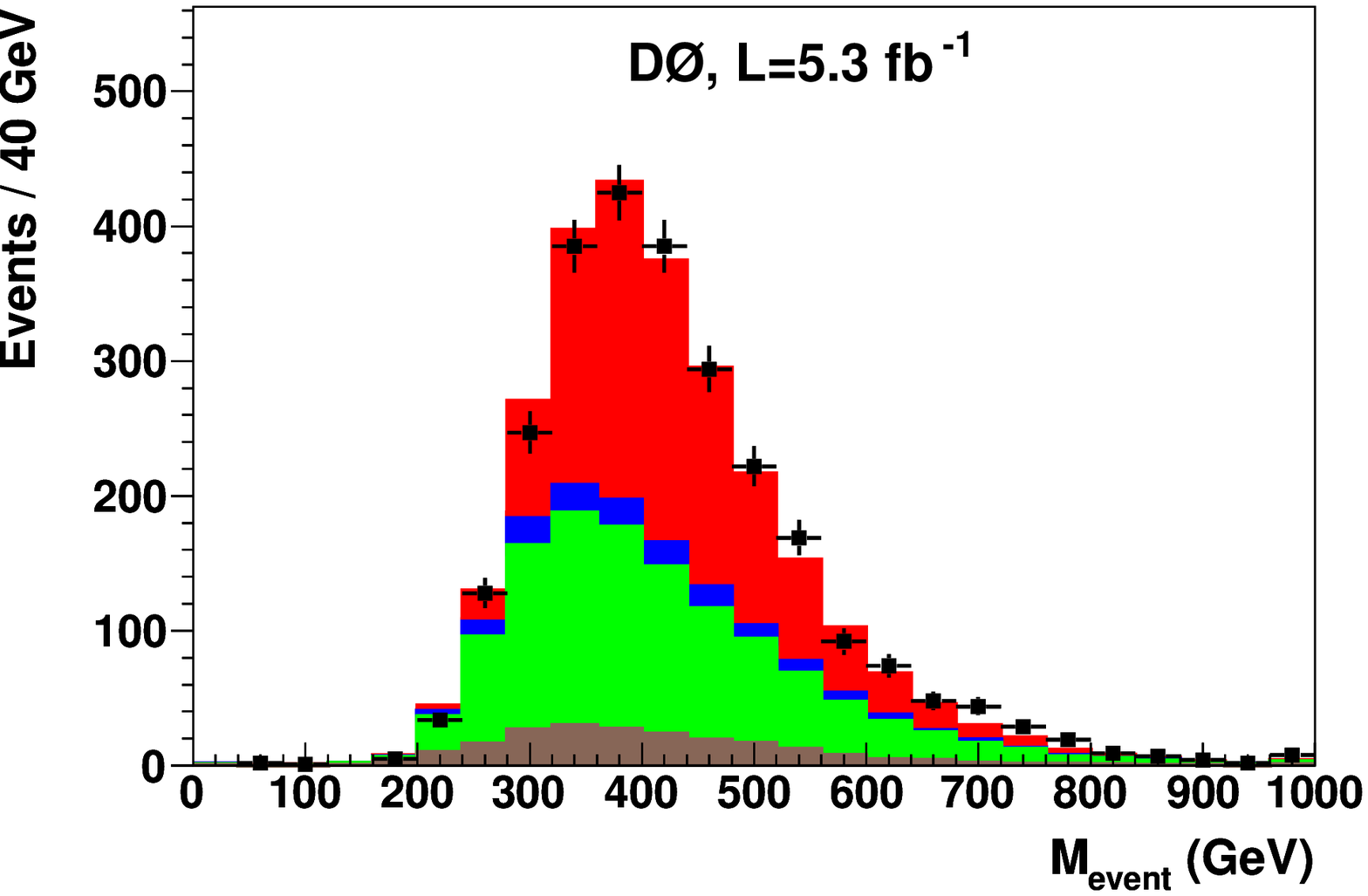}

\caption{(Color online) Distributions of input variables 
used in the RF discriminant for the $\ell$+$>3$\,jets channel 
in data overlaid with the predicted   
background and \ttb\ signal calculated using $\sigma_{t\bar{t}}=7.78$~pb 
as measured using the combined method.  
\label{fig:variables_control} }
\end{minipage}
\end{center}
\end{figure*}

\begin{figure*}[h]
\begin{center}
\begin{minipage}{18.0 cm}
\setlength{\unitlength}{1.0cm}
\begin{picture}(18.0,18.0)
\put(0.3,11.8){\includegraphics[width=8.6cm]{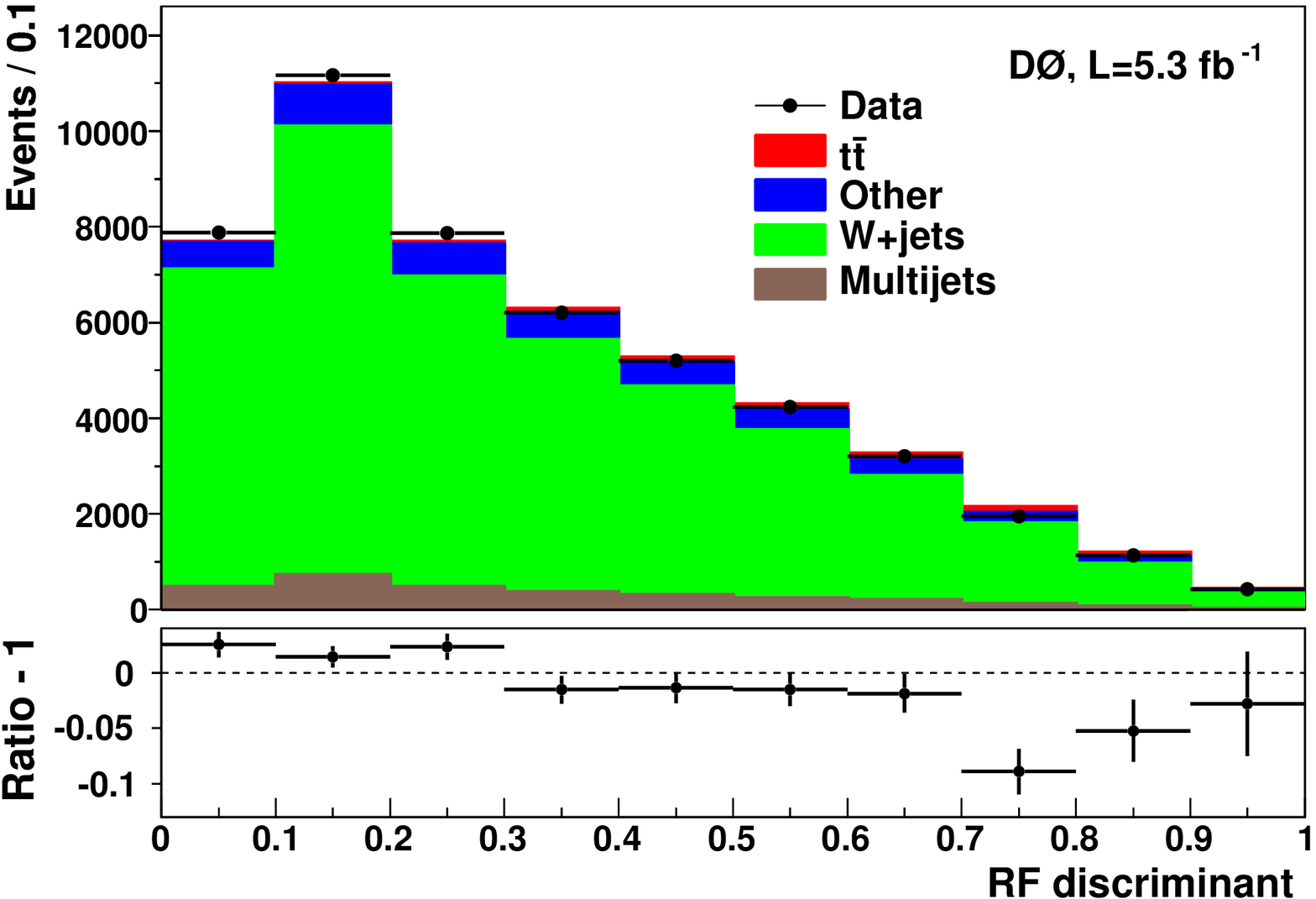}}
\put(9.0,11.8){\includegraphics[width=8.6cm]{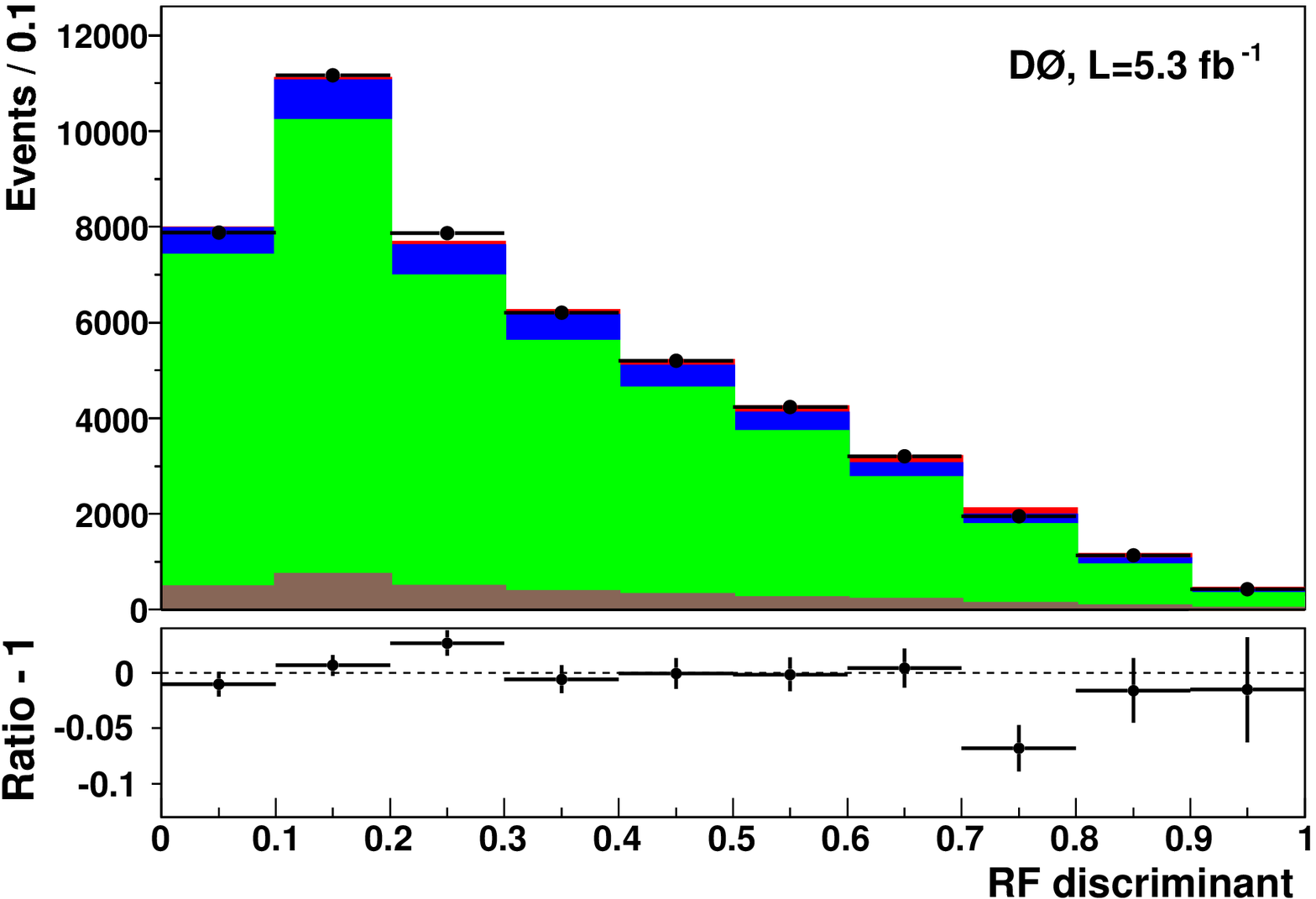}}
\put(0.3,5.9){\includegraphics[width=8.6cm]{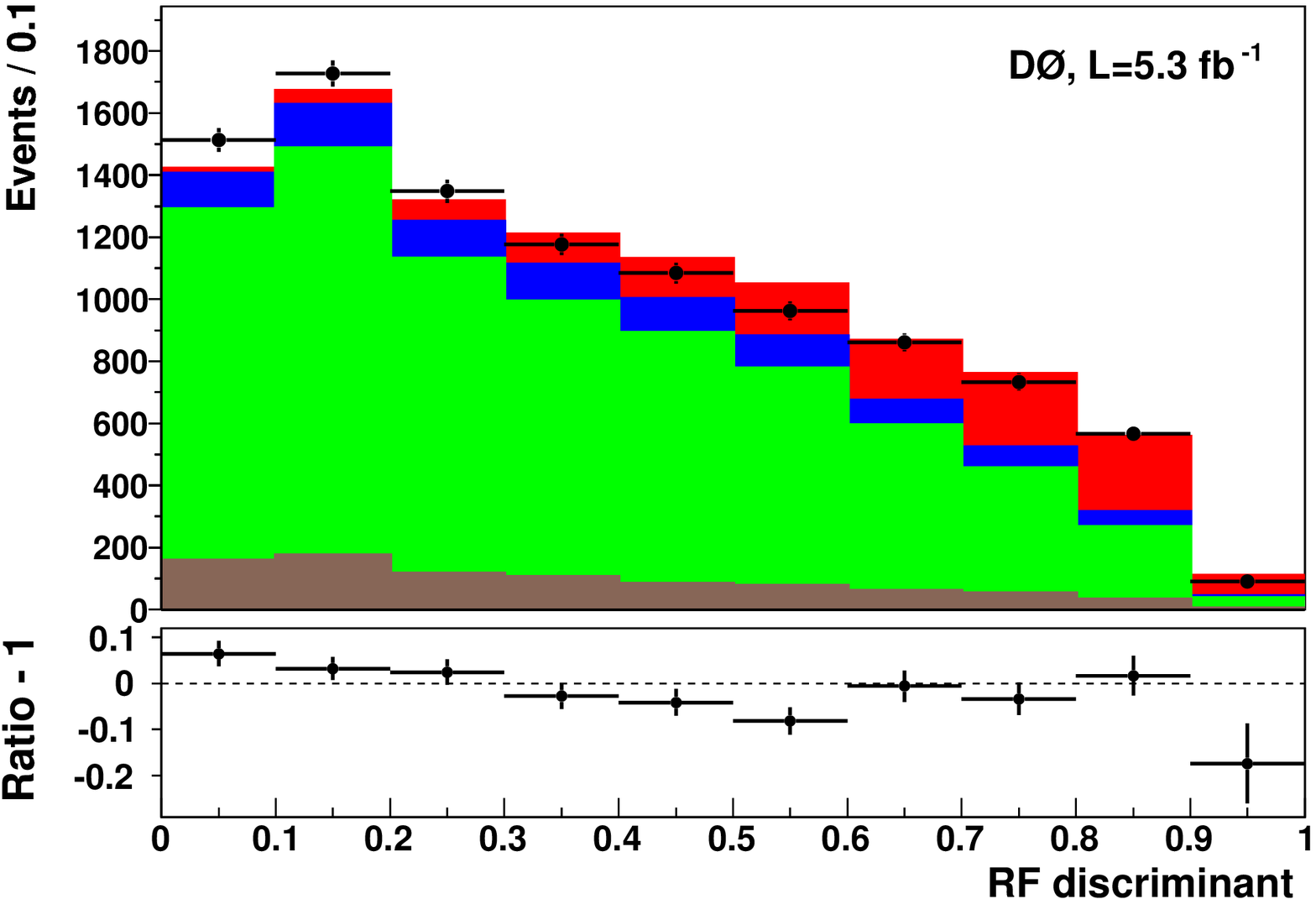}}
\put(9.0,5.9){\includegraphics[width=8.6cm]{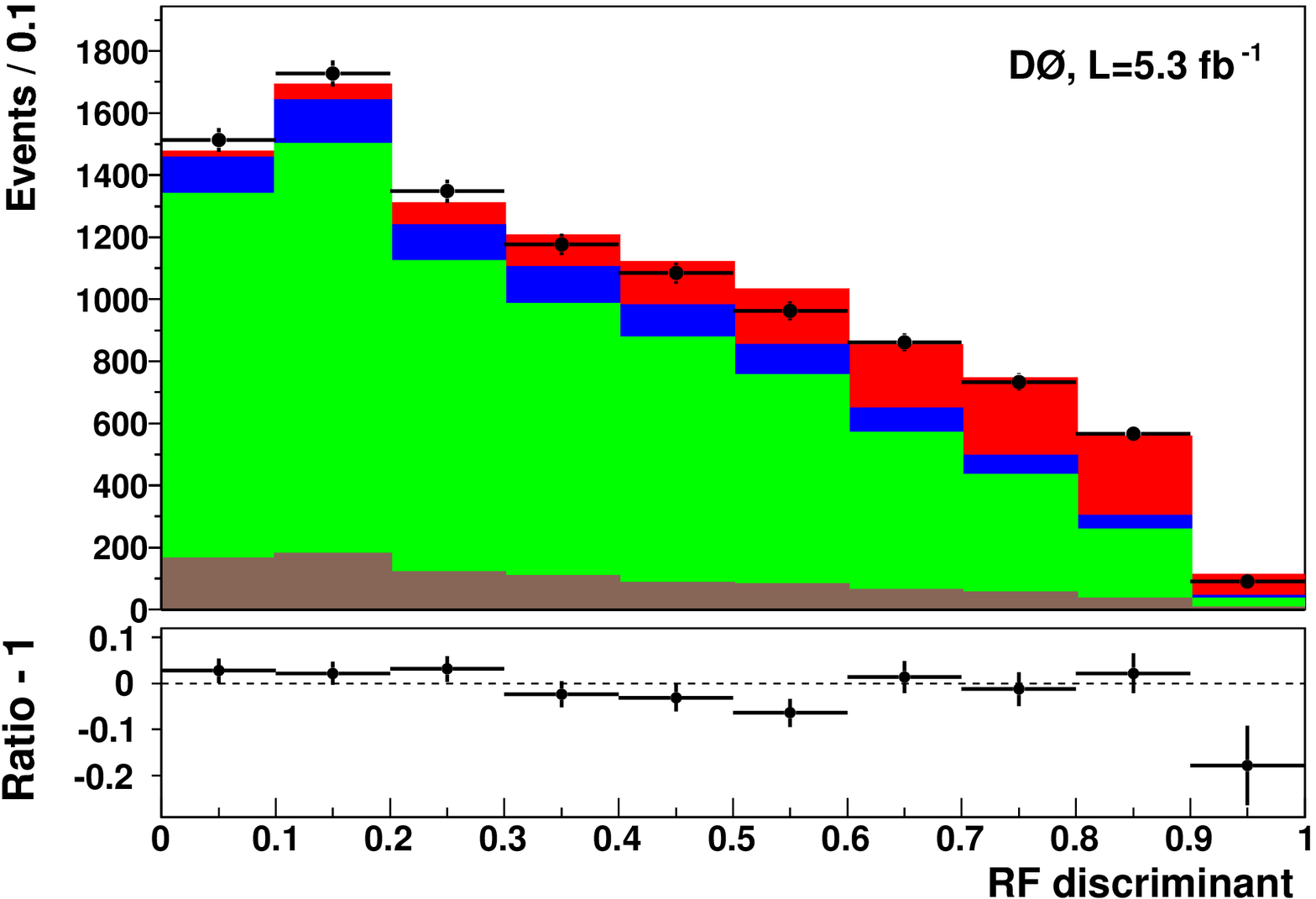}}
\put(0.3,0.0){\includegraphics[width=8.6cm]{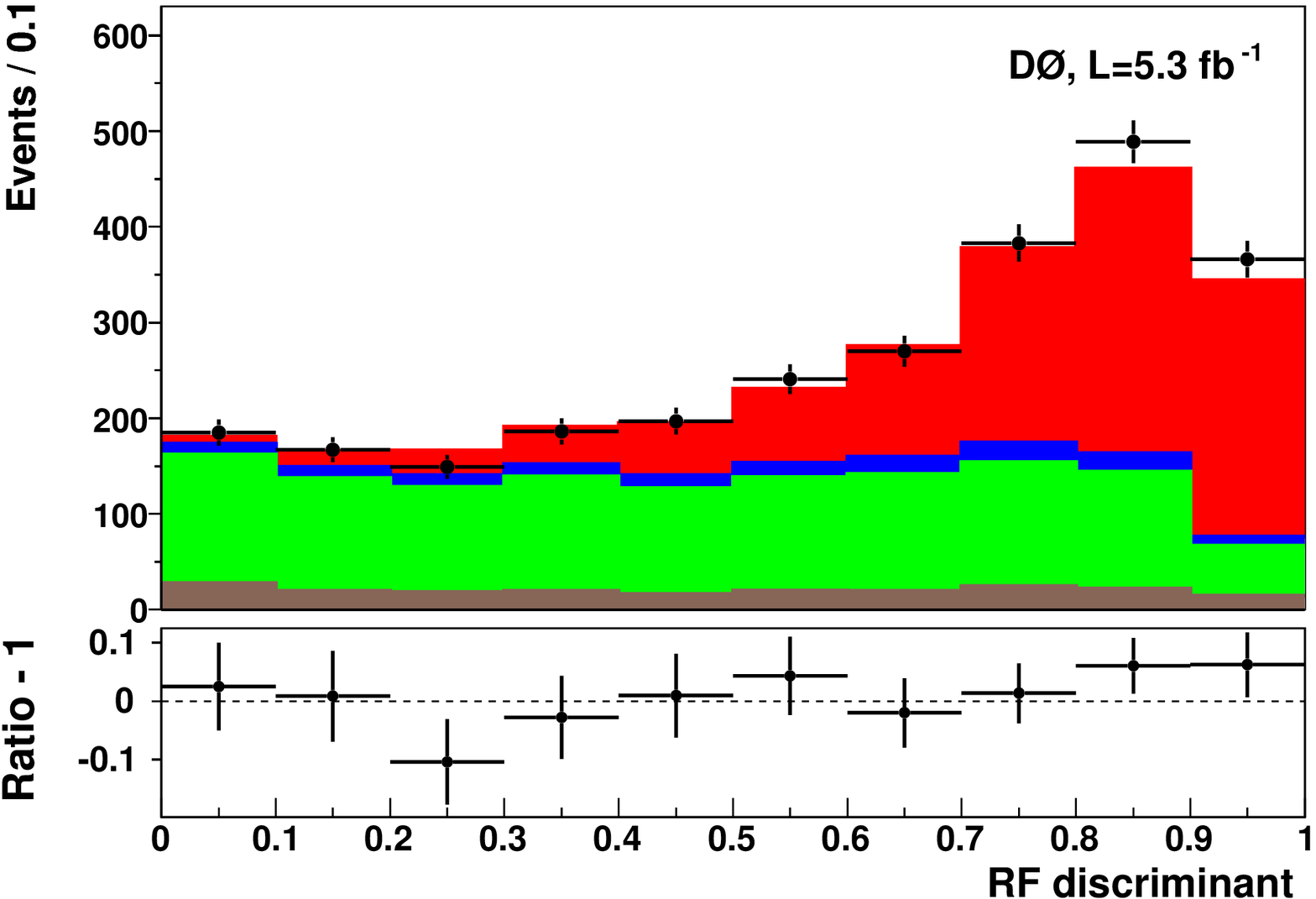}}
\put(9.0,0.0){\includegraphics[width=8.6cm]{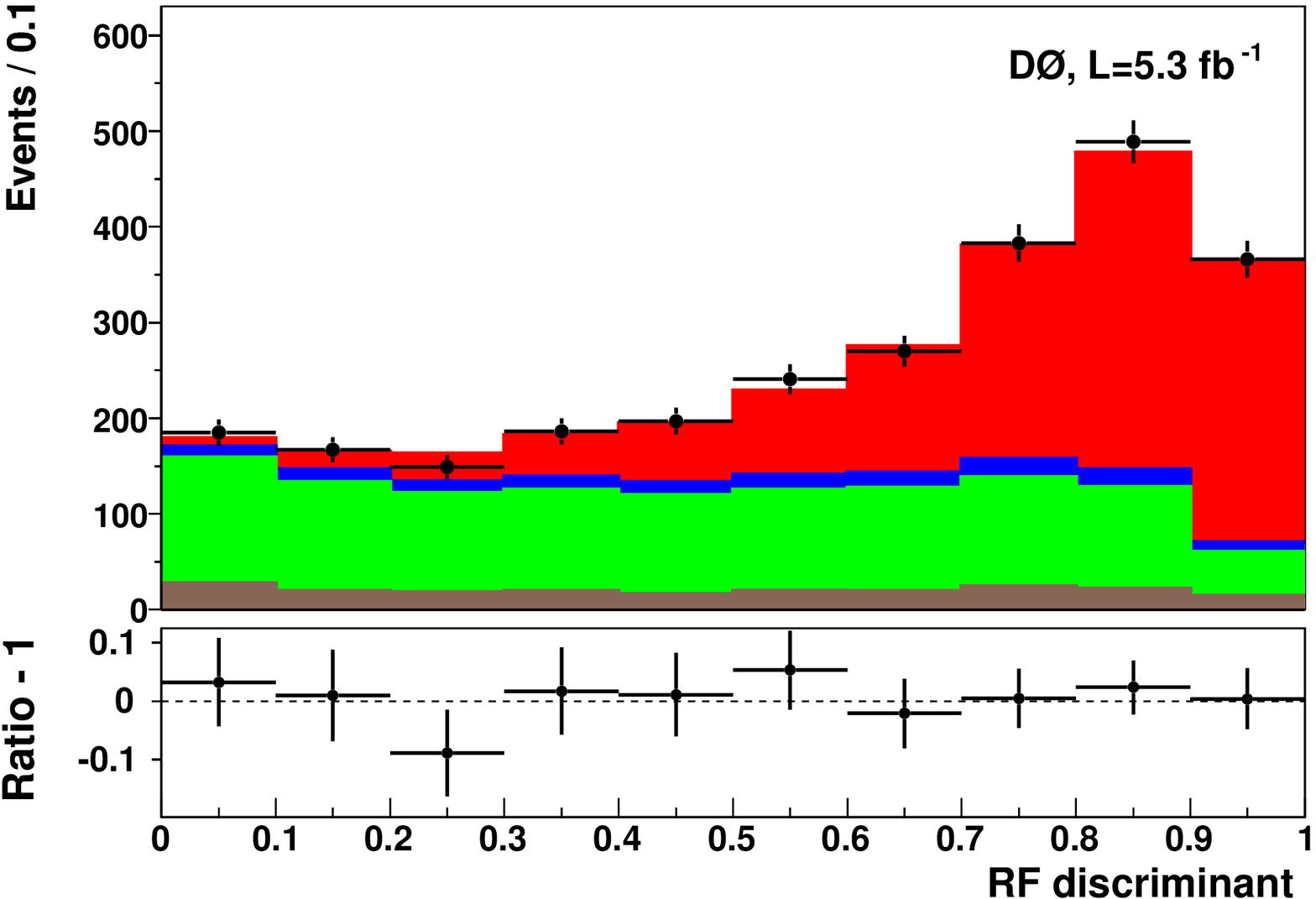}}
\put(1.5,5.18){(e)}
\put(10.2,5.18){(f)}
\put(1.5,11.08){(c)}
\put(10.2,11.08){(d)}
\put(1.5,16.98){(a)}
\put(10.2,16.98){(b)}
\end{picture}
\end{minipage}
\end{center}
\caption{(Color online) Output of the RF discriminant for 
(a) and (b) $\ell$+2\,jets,
(c) and (d) $\ell$+3\,jets and
(e) and (f) $\ell$+$>3$\,jets events, for backgrounds and a \ttbar\ signal based on 
the  cross section obtained with the kinematic method. The ratio of data over MC prediction is also shown.
The left plots (a, c, and e)  show the results with the nuisance
parameters fixed at value of zero. 
The right plots (b, d, and f) show the results when the nuisance parameters are determined
simultaneously with the $t\bar{t}$ cross section in the fit.
In the left and right plots the contribution
from the $t\bar{t}$ signal is normalized to the results of the cross section measurement,
$\sigma_{t\bar{t}}=7.00$ and $7.68$~pb, respectively
\label{fig:measured_discriminant_ljets_topo_p20} }
\end{figure*}

Figure~\ref{fig:variables_control} shows distributions 
for several of the input variables in the data compared to the 
sum of expected contributions from  
\ttb\ signal and backgrounds for the
$\ell$+$>3$ jets channel. 
The outputs of the RF discriminant are presented in
Fig.~\ref{fig:measured_discriminant_ljets_topo_p20} for the 
$\ell+2$, $\ell+3$ and $\ell$+$>3$ jets channels. 

Figure~\ref{fig:variables_control} indicates good agreement of data with expectation for 
$m_t=172.5$ GeV. 
Similar levels of agreement between data and prediction are 
observed in all other channels. The normalizations shown in Fig.~\ref{fig:measured_discriminant_ljets_topo_p20} are based on the results of the kinematic method.
 The distributions in Figs.~\ref{fig:measured_discriminant_ljets_topo_p20}(a, c, e) are the results when only fitting  \sigmatt; Figs.~\ref{fig:measured_discriminant_ljets_topo_p20}(b, d, f) 
show the result when the $t\bar{t}$ cross section is fitted together with other parameters, as shown in Eq.~\ref{eq:lhood} and  described in Sec.~\ref{sec:xs}.

\subsection{Cross Section Measurement}
\label{sec:xs}
To measure the $t\bar{t}$ cross section for the 
 kinematic analysis,  
we perform a binned maximum likelihood fit of the distributions in 
the RF discriminant to data.   
We use templates from MC 
for  dilepton and \ljets\ contributions to the \ttbar\ signal, as well as 
for $WW$, $WZ$, $ZZ$,
$Z$+jets, single top quark ($s$- and $t$-channel), and $W$+jets
backgrounds. 
The MJ template comes from data, and the amount of MJ
background is constrained within the uncertainties 
resulting from the matrix method.

We account for systematic uncertainties in the maximum likelihood fit by assigning a 
 parameter to each independent systematic variation. These ``nuisance'' parameters  are 
allowed to vary in the maximization of the likelihood function within uncertainties,  therefore the measured $t\bar{t}$ cross section
can be different from the value obtained if the
 parameters for the systematic uncertainties are not included in the fit.
The effects of a source of systematic uncertainty that is fully correlated
among several channels are controlled by a single parameter in these channels. 

The likelihood
function is defined as:
 \begin{eqnarray} \lefteqn{ {\cal L} =} \\ \nonumber
& & \prod_{j=1}^{12}{\left[\prod_i {\cal{P}}^{j}(n_i^{o},\mu_i)\right]{\cal{P}}^{j}(N_{LT}^{o},N_{LT})}  
 \prod_{k=1}^{K} {\cal G}(\nu_k;0,{\rm SD}) \,,\label{eq:lhood}\end{eqnarray}
where 
${\cal G}(\nu_k;0,{\rm SD})$ denotes the Gaussian probability density with mean at 
zero and width corresponding 
to one standard deviation (SD) of the considered systematic uncertainty,      
${\cal{P}}(n,\mu)$ denotes the Poisson probability density 
for observing $n$ events, given an expectation value of $\mu$,  
$N_{LT}$ denotes the number of events in the ``loose'' but not ``tight'' (``loose--tight'') 
sample required by the matrix method. The value of $N_{LT}$  is restricted
within Poisson statistics to the observed number of events, $N_{LT}^{o}$, 
in the ``loose--tight'' sample, 
ensuring the inclusion of the statistical uncertainty in the MJ
prediction.  
The first product runs over twelve data sets $j$ and all bins 
of the discriminant $i$; $n_i^{ o}$ is the content of bin $i$ 
in the selected data sample; and $\mu_i$ is the expectation for 
bin $i$. This expectation is the sum of the predicted
background and the expected number of \ttbar\ events, which depends on \sigmatt.  
The last product runs over all independent sources of systematic
uncertainties $k$, with $\nu_k$ being the corresponding nuisance parameters and  $K$ the total number of independent sources $k$.

Since the discriminant for the MJ background is not
determined from MC simulation but 
from the ``loose--tight'' data sample, it has a small contribution from events 
with leptons in the final state. 
This contamination of the MJ distribution is taken into account by using the 
corrected number of events expected in each bin of the discriminant functions used 
in Eq.~\ref{eq:lhood}:
\begin{eqnarray}
& &\mu_i(N_T^{ t\bar{t}},N_T^{ W},N_T^{ MC},N_T^{\rm MJ}) =  \\ \nonumber
& &\left(f_i^{ t\bar{t}}N_T^{t\bar{t}} + f_i^{ W}N_T^{W} +
\sum_{m}(f_i^{{\rm MC}_m}N_T^{{\rm MC}_m}) \right)\times \\ \nonumber
& & \times \left(1-\frac{\varepsilon_{b}}{1-\varepsilon_{b}}\frac{1-\varepsilon_{s}}{\varepsilon_{s}}\right)
+ \\ \nonumber
& + & f_i^{\rm MJ}\left(N_T^{\rm
MJ}+\frac{\varepsilon_{b}}{1-\varepsilon_{b}}\frac{1-\varepsilon_{s}}{\varepsilon_{s}}\left(N_T^{t\bar{t}}+N_T^{
W}+N_T^{\rm MC}\right)\right)\;, \nonumber
\end{eqnarray}
where $N_T^{t\bar{t}}$, $N_T^{ W}$, $N_T^{ MC}$, $N_T^{MJ}$ are the
numbers of \ttbar, ~$W$+jets, MC background (diboson, single top quark,
$Z$+jets) and MJ events in the tight lepton sample, index $m$ runs over all 
small backgrounds estimated from MC, and 
$f_i^{x}$ is the predicted fraction of contribution $x$ in bin $i$.

We minimize the negative of the log-likelihood function of Eq.~\ref{eq:lhood} 
as a function of \ttbar cross section and the nuisance parameters. The fit results for the \ttbar cross section and the nuisance parameters
are given by their values at the minimum of the negative log-likelihood function,
and their uncertainties are defined from the increase in the negative log-likelihood
by one-half of a unit relative to its minimum. Results of the fit are 
presented in Sec.~\ref{sec:results}.

\clearpage

\section{\boldmath ${\bm b}$-tagging Method} \label{sec:btagging}

\subsection{Discrimination}
The SM predicts that the top quark decays almost
exclusively into a $W$ boson and a $b$~quark ($t \rightarrow W b$).
Hence, besides using just kinematic information, the fraction of $t\bar{t}$ events
in the selected sample can be enhanced using $b$-jet identification.  
To measure the $t\bar{t}$ cross section, we use final states with exactly three jets and more than three jets 
and further separate each channel into events with $0$, $1$, and $>1$
$b$-tagged jets, obtaining 24 mutually exclusive data samples.

\subsection{Cross Section Measurement}
\begin{figure}[ht]
\begin{center}
\setlength{\unitlength}{1.0cm}
\begin{picture}(5.0,11.0)
\put(-1.5,0.0){\includegraphics[width=8.0cm]{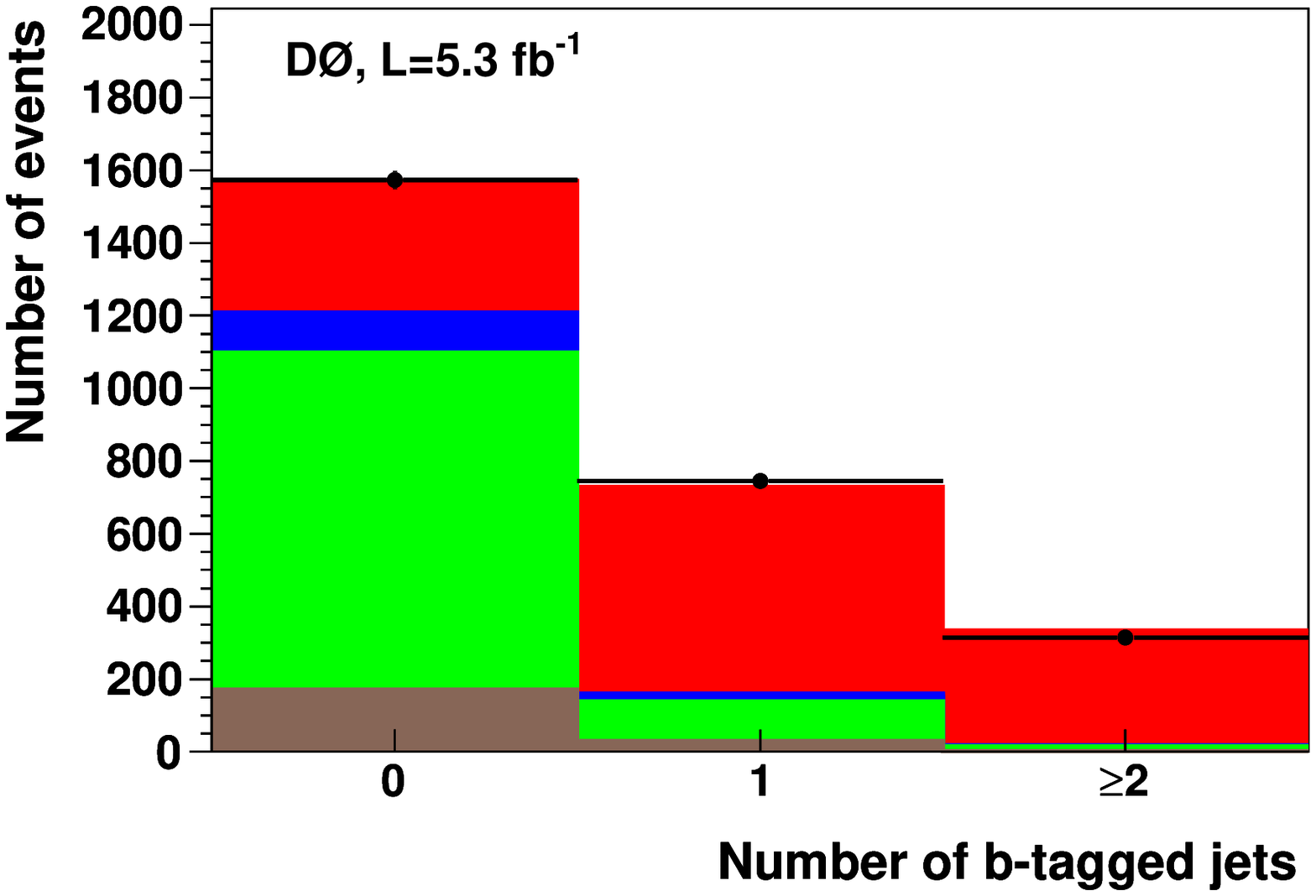}}
\put(-1.5,5.8){\includegraphics[width=8.0cm]{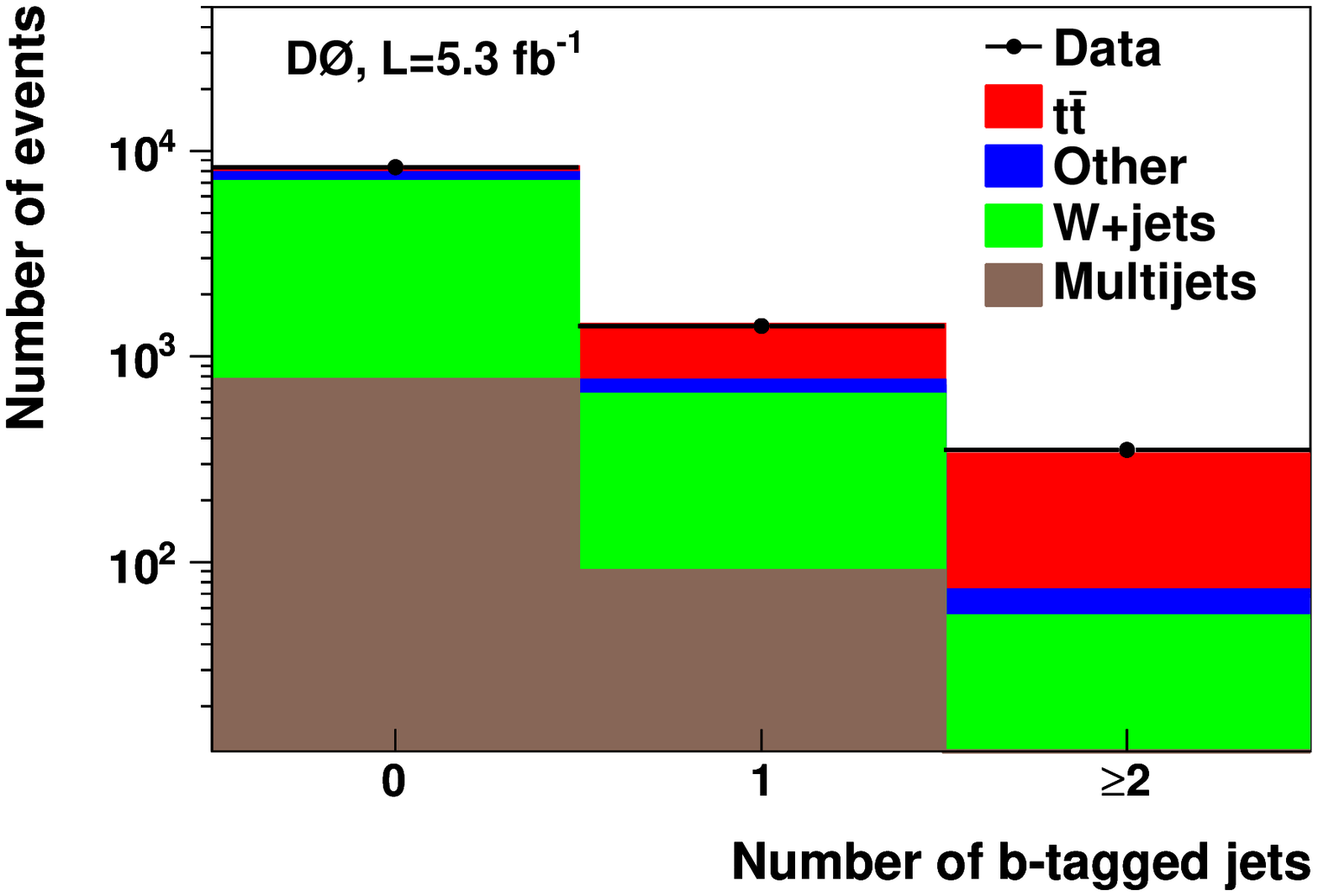}}
\put(4.95,2.5){(b)}
\put(4.95,8.6){(a)}
\end{picture}
\end{center}
\caption{(Color online) Distributions of events with $0$, $1$, and $>1$ $b$-tagged
jets for (a) $\ell$+3\,jets and (b) 
$\ell$+$>3$\,jets, for backgrounds and contributions from \ttbar\ signal for 
 $\sigma_{t\bar{t}}=8.13$\, pb as  measured  using the $b$-tagging method.  
\label{fig:tagbin_btag}}
\end{figure}

As discussed is Sec.~\ref{sec:samplecompo}, before applying 
$b$-tagging, the contribution from the $W$+jets background is normalized to the difference between
data and the sum of \ttbar signal and all other sources of background.
Since the $W$+jets background normalization depends on 
the $t\bar{t}$ cross section, the measurement of the cross section and
the $W$+jets normalization determination are performed simultaneously. 
Details of this method, as well as the
general treatment of systematic uncertainties are described in Ref.~\cite{prdbtag}. 
The fit of the $t\bar{t}$ cross section to data is
performed using a binned maximum likelihood fit for the predicted number of
events, which depends on $\sigma_{t\bar{t}}$. 
The likelihood is defined as a  
product of Poisson probabilities for all 24 channels $j$:  
 \begin{eqnarray} \lefteqn{ {\cal L} = }\\ \nonumber
& & \prod_{j=1}^{24} {\cal P}(n_{j},\mu_{j})  {\cal{P}}^{j}(N_{LT}^{o},N_{LT}) \prod_{k=1}^{K}
 {\cal G}(\nu_k;0,{\rm SD}) \,,\label{eq:poisson1}\end{eqnarray}
and systematic uncertainties are 
incorporated into the fit in the same way as described in Sec. \ref{sec:xs}.  
Figure \ref{fig:tagbin_btag} shows the distributions of events with  $0$, $1$, and $>1$ $b$-tagged 
jets for events with three and more than three jets in data compared to the sum of 
predicted background and measured \ttbar signal using $b$-tagging method. 
Results for this method are given in Sec.~\ref{sec:results}. 

\clearpage

\section{Combined Method} \label{sec:combi} 
In the combined method, kinematic information and $b$-jet
       identification are used.
We split the selected sample into events with $2$, $3$, and
$> 3$ jets and into $0$, $1$, and $> 1$ $b$-tagged jets and construct 
RF discriminant functions as described in Sec.\ \ref{sec:topoxsec}
for the channels dominated by the background.  

For events with $> 2$ jets but no $b$-tagged jet, we construct
a RF discriminant using the same six 
variables as for the kinematic method described in Sec.~\ref{sec:topoxsec}. 
For events with three jets and one $b$-tag, we construct
discriminants using only ${\mathcal A}$, $\it{ H_T^3}$ and
$\it{M_T^{j_2\nu \ell}}$. 
For all other subchannels, we do not form RF discriminants,  
but use the $b$-tagging method described in
Sec.~\ref{sec:btagging}. The signal purity is
already high in those channels except for the ones with two jets, which
do not have a sizable signal contribution and are used to measure 
the $W$+jets heavy-flavor scale factor $f_H$ which is the source of one of the largest uncertainties in the $b$-tagging analysis. 

To reduce this source of uncertainty, we 
measure $f_H$ simultaneously with \sigmatt, assuming that 
$f_H$ for $Wb\bar{b}$ production is the 
same as for $Wc\bar{c}$ production and that it does not depend on 
the number of jets in the event. Since 
sources of uncertainty such as light-flavor jet tagging rates  
are correlated with the value of $f_H$, 
and  in turn, $f_H$ is anti-correlated with the \ttbar\ cross section,
the total uncertainty on the measured \sigmatt decreases.  
The main constraint on $f_H$ is provided  
by the $2$-jets channels with $0$, $1$,
and $>1$ $b$-tagged jets. For this reason the RF discriminant was not used for the $2$-jets channels 
in contrast to the measurement using only kinematic information  
(Sec.~\ref{sec:topoxsec}). 

The cross section is measured using the likelihood function of
Eq.~\ref{eq:lhood} for channels where a RF discriminant
is calculated, and using Eq.~\ref{eq:poisson1} for all other
channels where the $b$-tagging method is performed. 
In the minimization procedure, we multiply appropriate likelihood functions 
for each channel and perform a fit to data assuming the same 
\ttbar cross section for all considered channels.  
Systematic uncertainties for each channel are
incorporated  as described in Sec.~\ref{sec:xs}. The $W$+jets 
heavy-flavor scale factor enters the calculation of  
the predicted number of $W$+jets events,  
 $N(W) \propto N(W+{\rm lf}) + f_H N(W+{\rm hf})+f_{Wc}N(W+c)$, where  
$f_{Wc}$ denotes the scale factor needed for $W+c$ events.  
 A change in $f_H$ 
results in a change in the predicted number of $W$+jets events
in each tag category without changing the total number of $W$+jets 
events in the sample prior to applying the $b$-tagging requirement which is   normalized to data. 

\begin{figure}[ht]
\begin{center}
\setlength{\unitlength}{1.0cm}
\begin{picture}(5.0,17.0)
\put(-1.5,11.6){\includegraphics[width=8.0cm]{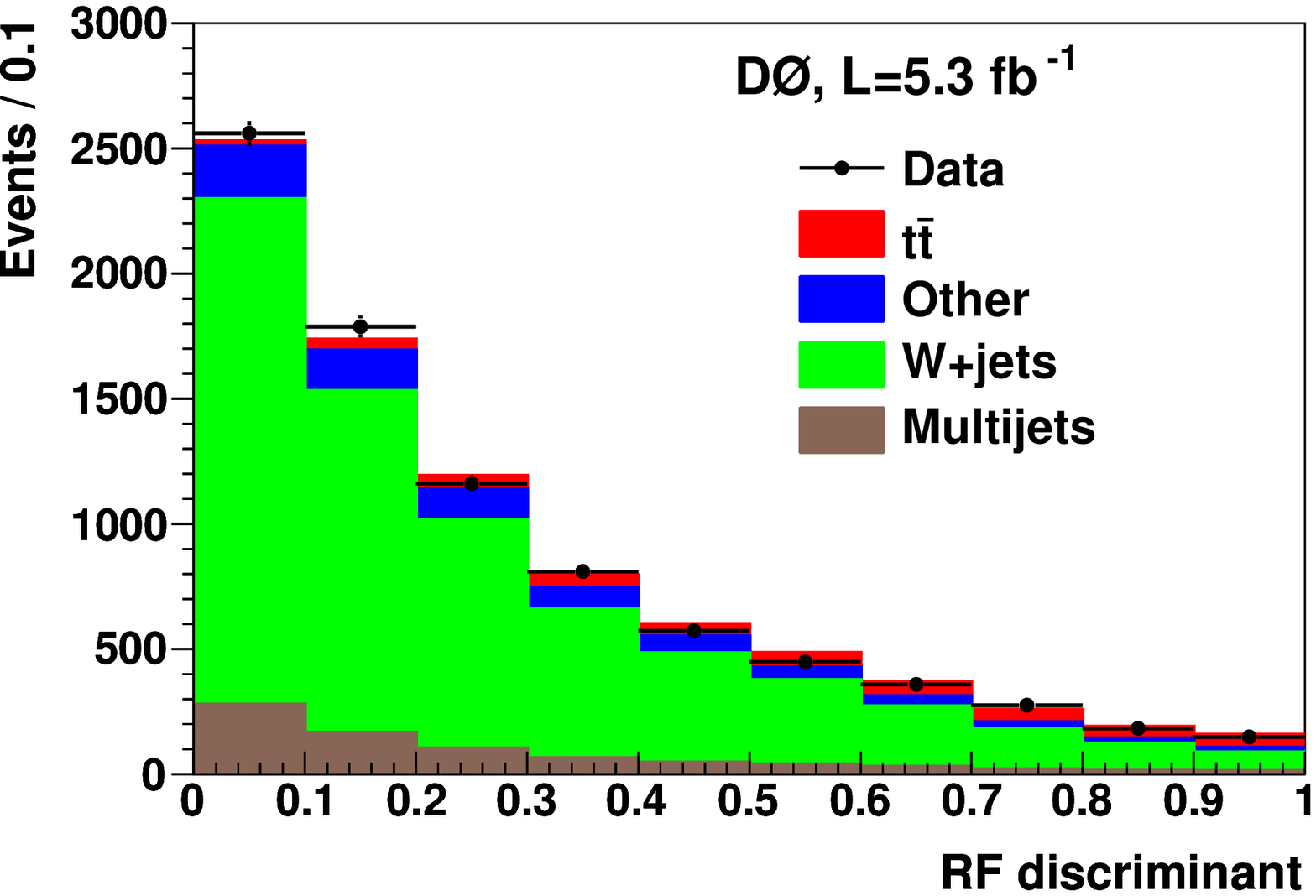}}
\put(-1.5,5.8){\includegraphics[width=8.0cm]{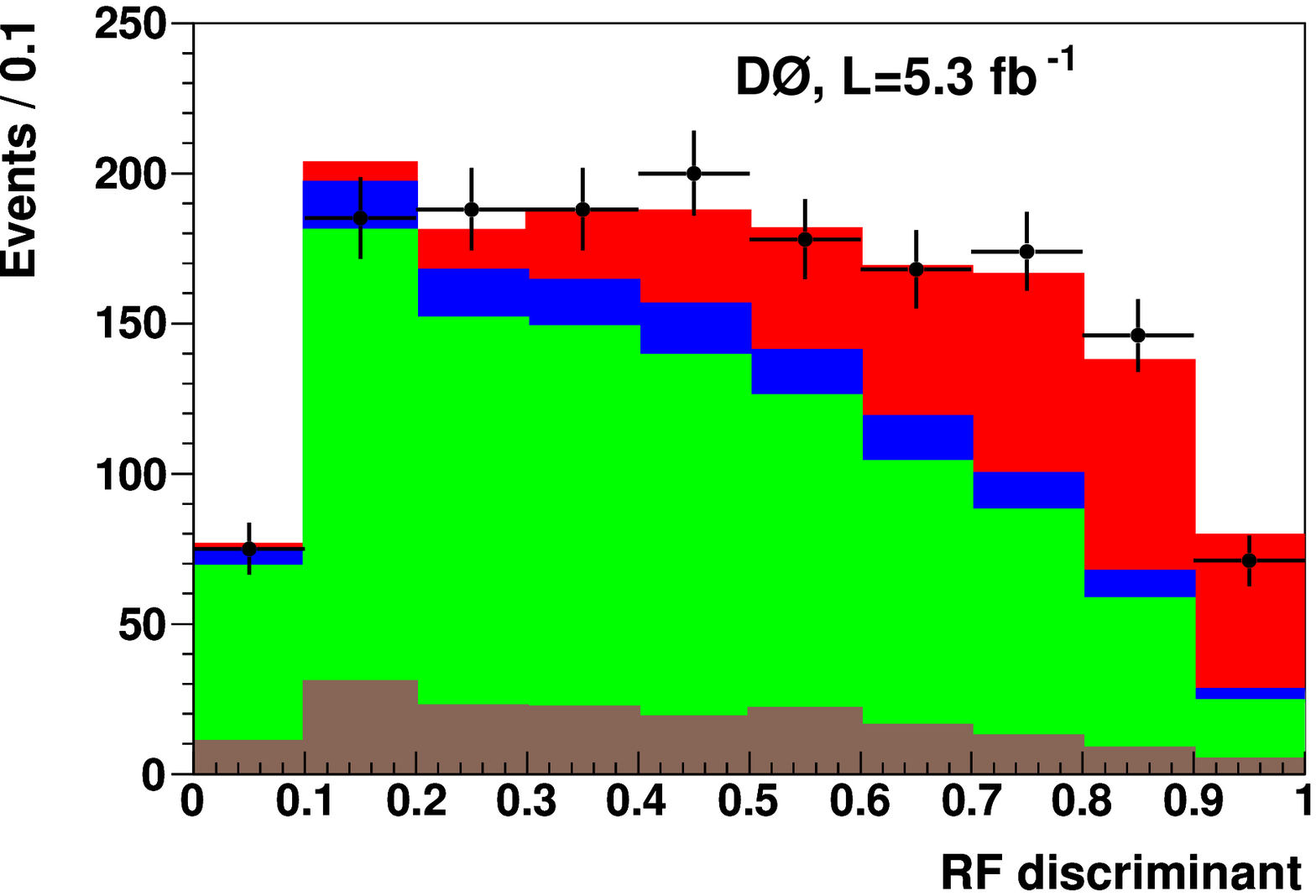}}
\put(-1.5,0.0){\includegraphics[width=8.0cm]{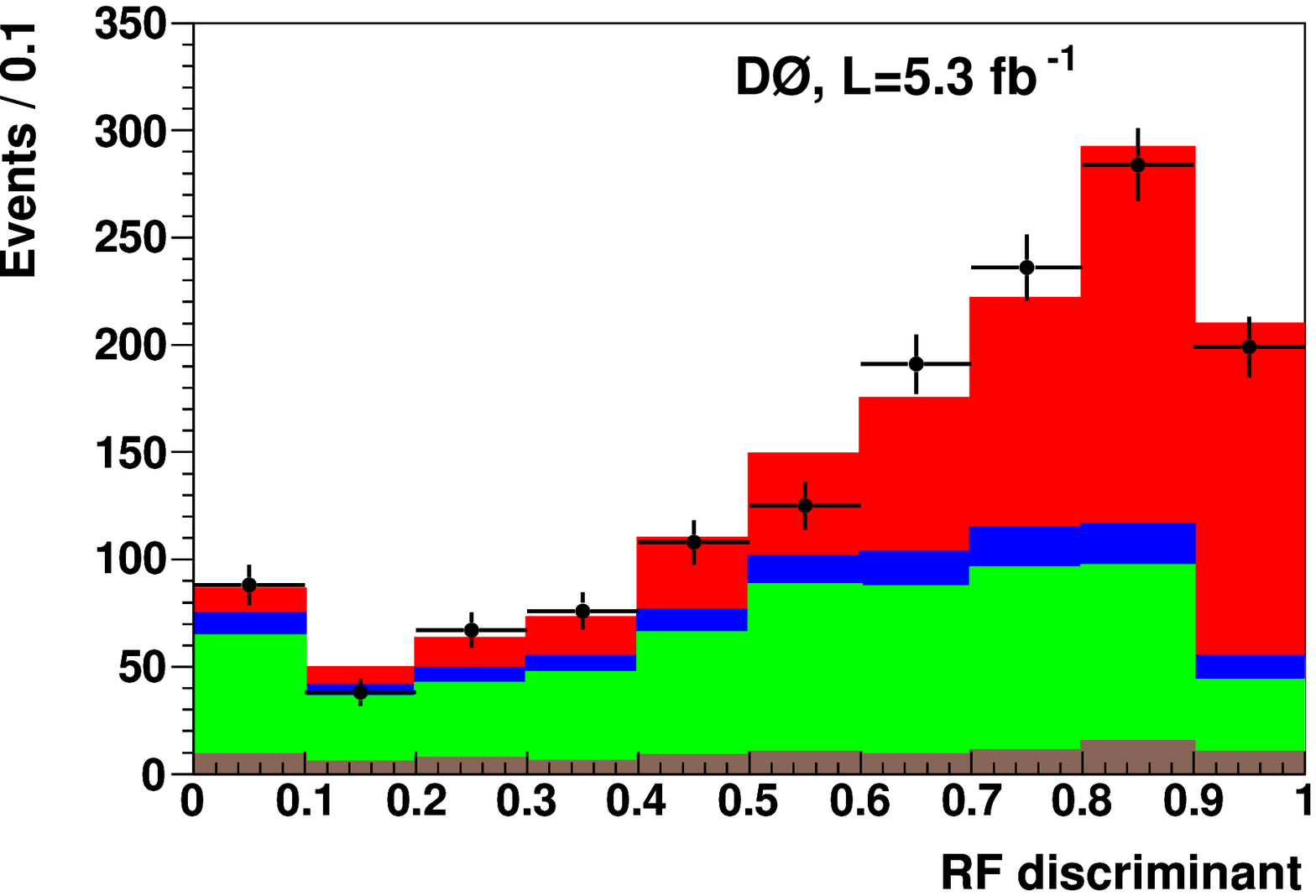}}
\put(-0.15,4.8){(c)}
\put(-0.15,10.6){(b)}
\put(-0.15,16.4){(a)}
\end{picture}
\end{center}
\caption{(Color online) Output of the RF discriminant for (a) $\ell$+3\,jets,  
(b) $\ell$+$>3$\,jets for events without $b$-tagged jets, and
(c) $\ell$+3\,jets with one $b$-tagged jet, for  
backgrounds and contributions from \ttbar\ signal for a 
cross section of $7.78$~pb as measured with the combined method.
\label{fig:measured_discriminant_ljets_comb} }
\end{figure}

\begin{figure}[ht]
\begin{center}
\setlength{\unitlength}{1.0cm}
\begin{picture}(5.0,17.0)
\put(-1.5,11.6){\includegraphics[width=8.0cm]{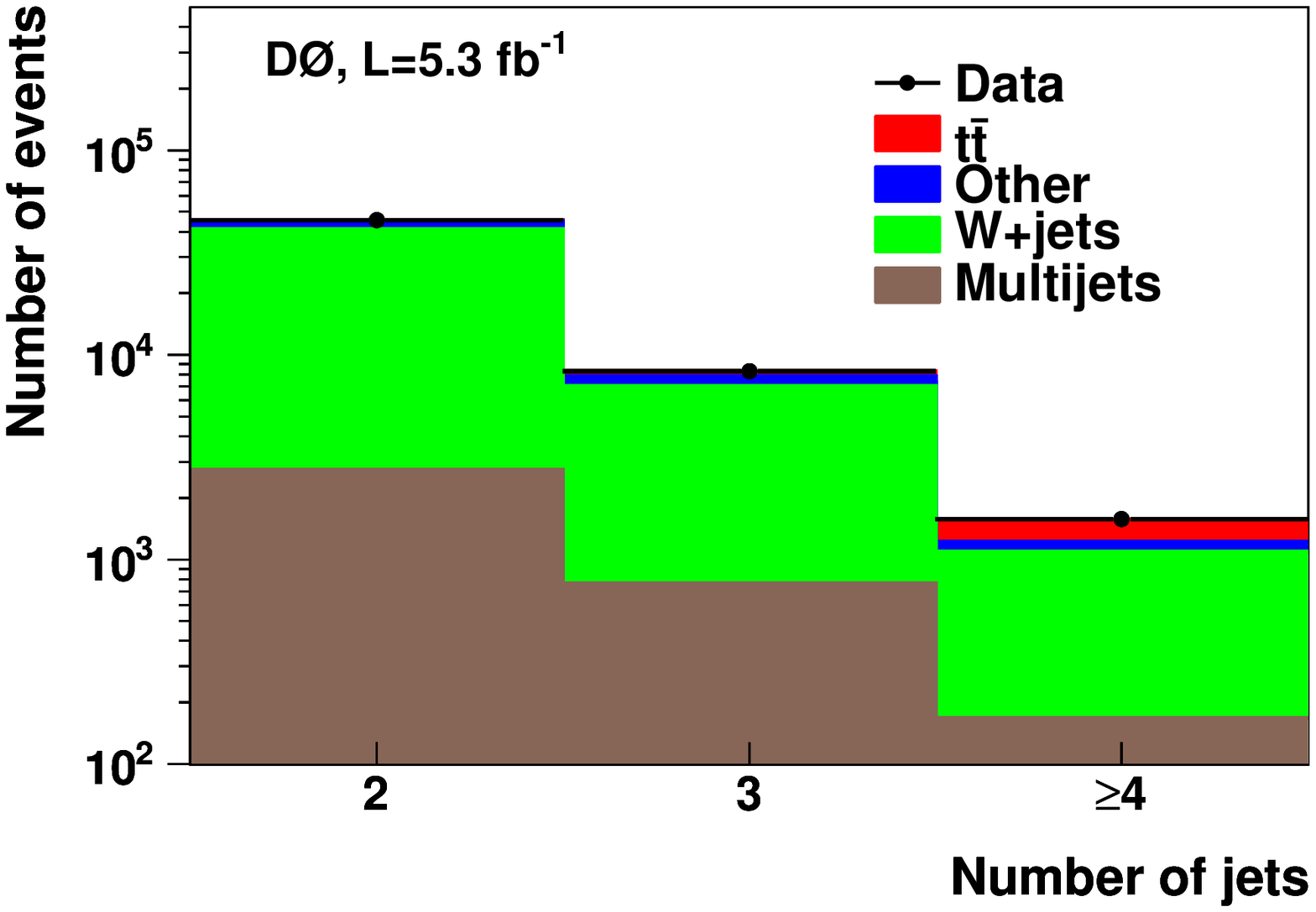}}
\put(-1.5,5.8){\includegraphics[width=8.0cm]{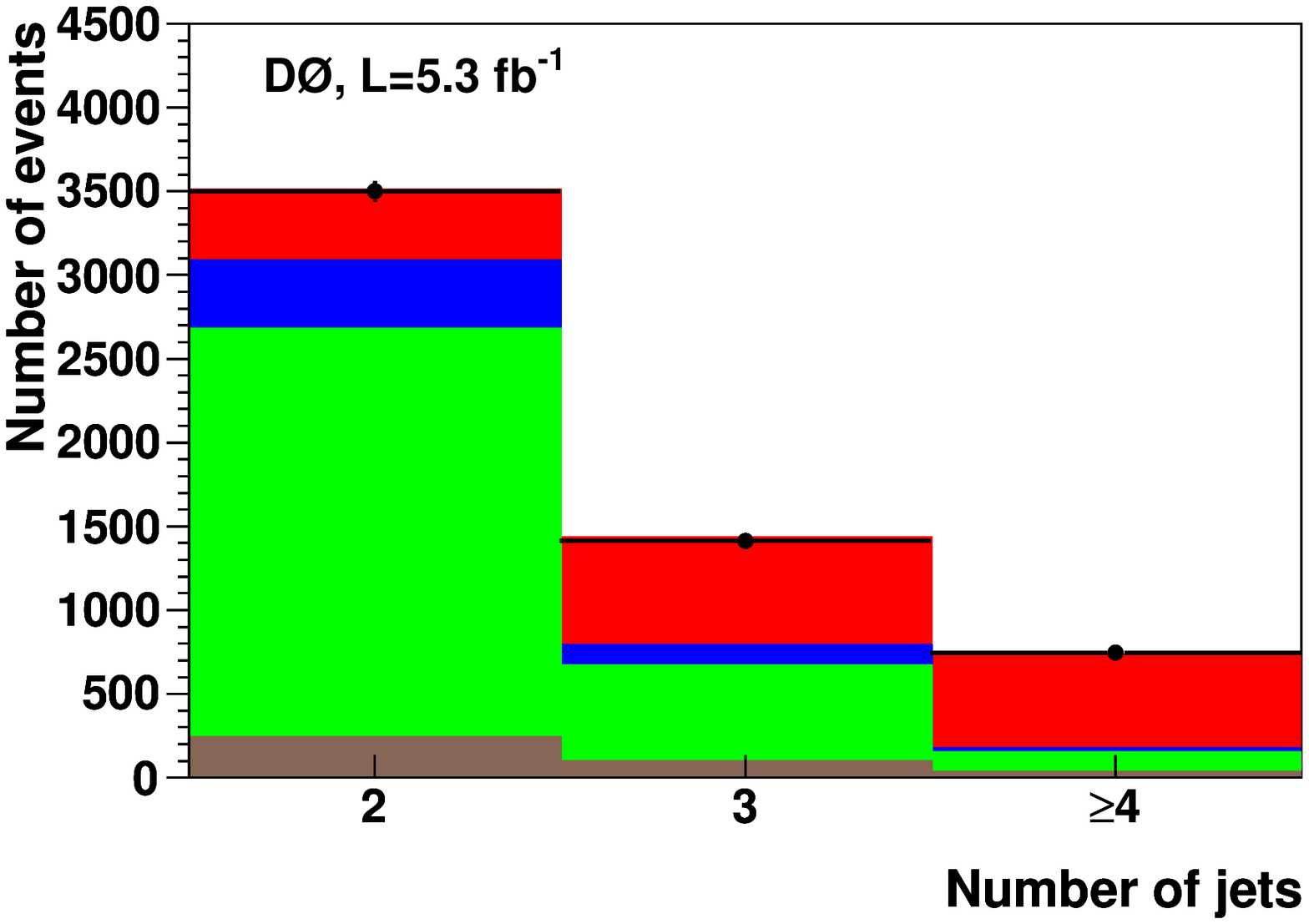}}
\put(-1.5,0.0){\includegraphics[width=8.0cm]{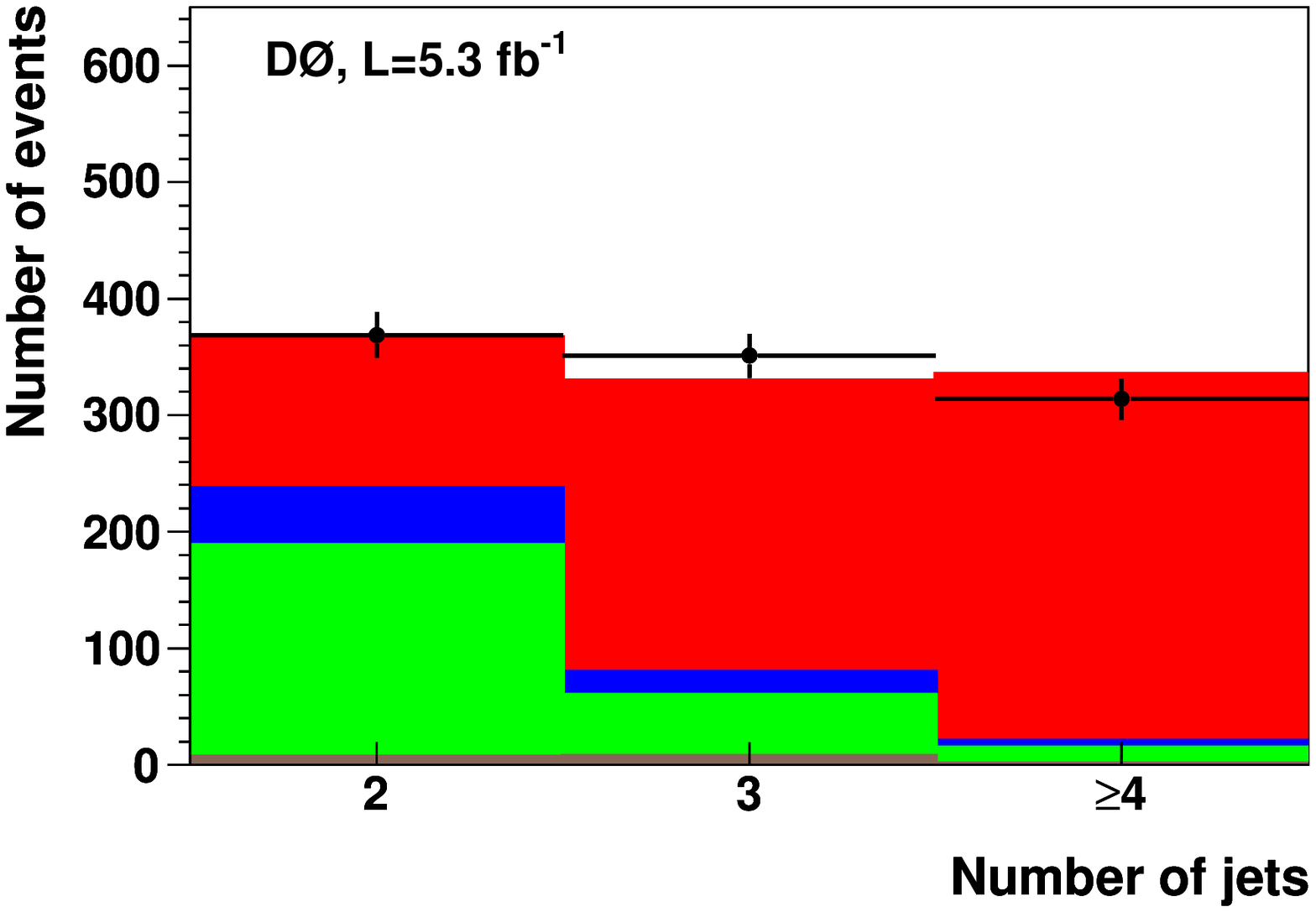}}
\put(4.95,4.3){(c)}
\put(4.95,10.1){(b)}
\put(4.95,15.9){(a)}
\end{picture}
\end{center}
\caption{(Color online) Jet multiplicity distributions for events 
with (a) 0, (b) 1, and (c) $>1$ $b$-tagged jets for backgrounds and 
contributions from \ttbar\ signal  for a 
cross section of $7.78$~pb as measured with the combined method.
\label{fig:tagbin_combi_l4jets}}
\end{figure}

Figure~\ref{fig:measured_discriminant_ljets_comb} shows the distribution of the RF 
discriminant for the $\ell$+3\,jets and
$\ell$+$>3$\,jets channels containing no $b$-tagged jets and for the 
$\ell$+3\,jets channel containing one $b$-tagged
jet. Figure~\ref{fig:tagbin_combi_l4jets} shows distributions of the
number of jets for events with different numbers of $b$-tagged jets. 
In both figures we use the measured values of \sigmatt and $f_H$ (see Sec.~\ref{sec:results}) 
as well as the nuisance parameters obtained from the fit.

\section{Systematic uncertainties} \label{sec:systs}
Different sources of
systematic uncertainty can affect selection efficiencies, $b$-tagging
probabilities, and the distributions of the RF discriminants.  
The sources that affect the selection efficiencies are 
electron and muon identification efficiencies, electron and muon
trigger efficiencies, modeling of additional \ppbar\ collisions in the
MC simulation, 
corrections on the longitudinal distribution of the PV in the MC simulation 
and data-quality requirements (summarized under ``other'' in the tables of uncertainties), 
uncertainties on the  normalization of the background obtained using  MC, and  
uncertainties on the modeling of the signal. 

The uncertainties due to $b$-tagging include corrections to the 
$b$, $c$, and light-flavor jet tagging rates, the track multiplicity requirements on jets which are candidates for $b$-tagging (called ``taggability''), and on the possible differences in the 
calorimeter response between  $b$ jets and light flavor jets. In addition, 
uncertainties in 
selection efficiencies and $b$-tagging probabilities can arise 
from limited statistics of MC samples and from the modeling of \ttbar 
signal. The latter includes PDF uncertainty,  
the difference between tuning of $b$-fragmentation to LEP or
SLD data~\cite{bfragtechninote}, the difference between  
simulations 
using \alpgen\ or \mcatnlo~\cite{mcnlo}, and between \pythia\ or 
\herwig~\cite{herwig} for parton evolution and hadronization,
 and the uncertainties on modeling color re-connections and on calculating
initial and final state radiation. The uncertainty on the PDF is 
estimated by evaluating
the effect of 20 independent uncertainty PDF sets of
CTEQ6.1M~\cite{cteq61m} on the selection efficiency 
and $b$-tagging probabilities, and adding the resulting uncertainties 
in quadrature. 

The uncertainties on the MJ background obtained from
the matrix method include systematic uncertainties on
 $\epsilon_{s}$ and $\epsilon_{b}$ as well as statistical
uncertainties due to the limited size of the samples
used to model MJ background. Uncertainties 
on the flavor composition of $W$+jets and $Z$+jets processes are also 
taken into account.  

Uncertainties on the jet energy scale~\cite{jes} (JES) and  jet reconstruction 
and identification efficiencies affect the selection and $b$-tagging
efficiencies, and the    
discriminant distributions. The discriminant distributions are also affected by 
the limited statistics used to form the templates.
In the combined method, systematic
uncertainties that affect the discriminant distributions include
taggability and tagging rates for $b$, $c$, and light-flavor jets. 
The uncertainty on the integrated luminosity is 6.1\% ~\cite{d0lumi}, affecting the 
estimates of signal and background yields obtained from simulation.  

Jet energy scale, jet
energy resolution, and jet reconstruction and identification uncertainties 
have a large effect on the discriminant distributions for $W$+jets 
background and as a result, a large effect on the measured \sigmatt. 
Their influence can be reduced by including events with 
two jets, dominated by the $W$+jets background, in the fit. 
Due to the correlation of the considered systematic
uncertainties between the different channels, the corresponding
nuisance parameters are constrained by the background-dominated
two-jet channels, and affect the result mostly through
the samples with more jets, where the \ttbar\ content is higher.

The cross section fit with a simultaneous extraction
of the nuisance parameters also results in a better agreement
between data and the signal plus background prediction for
the discriminant distribution in background
dominated samples. An example of this effect is illustrated
in Fig.~\ref{fig:measured_discriminant_ljets_topo_p20}, where we perform a comparison of data and
the total signal plus background prediction for the case
in which only the $t\bar{t}$ cross section is a free parameter
of the fit and for the case in which also the nuisance
parameters are determined from the fit. Improvements can be seen when the additional parameters associated with systematic contributions are varied.

We take into account all
correlations between channels and run periods. All uncertainties are
taken as correlated between the channels except for contributions from
MC statistics, trigger efficiencies, and the isolated lepton and
fake rate required by the matrix method. 
Systematics uncertainties measured using independent Run~IIa and Run~IIb 
data sets and  are dominated by the limited statistics of these data sets are taken as uncorrelated, 
including trigger efficiencies, jet energy scale, jet
identification, jet 
energy resolution, taggability, and lepton identification.

\section{Results} \label{sec:results}
We quote the results for the $t\bar{t}$ cross section measurements using
the three different methods described above, assuming a value of the top
quark mass of $172.5$~GeV. In Sec.~\ref{sec:massdep} we discuss the dependence of the cross section measurement on the assumed value of the top quark mass.

\subsection{Kinematic method}

\label{sec:topores}
Table~\ref{tab:ljets_separate_channels_topo_nuisance} shows the
measured cross section in the \eplus and the \muplus\ channels, and for the combined \lplus\
channel for the kinematic method. 
Table~\ref{tab:ljets_syst_topo_nuisance} lists the
corresponding uncertainties. For each category of systematic uncertainties listed in Table~\ref{tab:ljets_syst_topo_nuisance}, only the corresponding nuisance parameters are allowed to vary. The column ``Offset'' shows the absolute shift of the measured $t\bar{t}$ cross section with respect to the result obtained including only statistical
uncertainties.  The columns ``$+\sigma$'' and ``$-\sigma$'' list the systematic uncertainty on the measured cross section for each category. 
For the ``fit result'' all nuisance parameters are allowed to vary at the same time, which can result in a different ``offset'' and different uncertainties on the final $t\bar{t}$ cross section than expected from a sum of the individual ``offsets'' and systematic uncertainties. The uncertainty given in the row ``fit result'' refers to the full statistical plus systematic uncertainty.

In the final fit, all nuisance parameters vary by less than one
SD from their mean value of zero. This also applies for the two other
methods used for the extraction of the cross section.

\begin{table}[h]
\newcommand\T{\rule{0pt}{2.6ex}}
\newcommand\B{\rule[-1.2ex]{0pt}{0pt}}
\begin{center}
\caption{Measured $t\bar{t}$ cross section using 
the kinematic method for separate and combined \lplus\ channels. 
The first quoted uncertainty denotes the statistical, the second the systematic contribution. The statistical uncertainty is scaled
from the statistical only $\sigma_{t\bar{t}}$ result in Table~\ref{tab:ljets_syst_topo_nuisance} to the final $\sigma_{t\bar{t}}$. 
The total uncertainty corresponds to the one in the row ``Fit result'' in Table~\ref{tab:ljets_syst_topo_nuisance}.
\label{tab:ljets_separate_channels_topo_nuisance} }
\begin{tabular}{lc  c  c} \hline \hline
Channel & $e$+jets & $\mu$+jets & $\ell$+jets \\[1pt]\hline \\[-7pt]
$\sigma_{t\bar{t}}$[pb] & $6.87\pm0.37^{+0.72}_{-0.52}$ & $8.04\pm0.48^{+0.75}_{-0.59}$ & $ 7.68\pm0.31^{+0.64}_{-0.56}$\\[4pt]
\hline\hline
\end{tabular}
\end{center}
\end{table}

\begin{table*}[ht]
\newcommand\T{\rule{0pt}{2.6ex}}
\newcommand\B{\rule[-1.2ex]{0pt}{0pt}}
\begin{center} 
\begin{minipage}{5.5 in}
 \caption{Measured $t\bar{t}$ cross section and the breakdown of uncertainties 
 for the kinematic method in the \lplus\ channel. 
 The offsets show how the mean value of the measured
 cross section is shifted due to each source of systematic uncertainty. In each line, all but the considered source of systematic
   uncertainty are ignored. 
The $\pm \sigma$ give the impact on the measured
 cross section when the nuisance parameters describing the considered category are 
 changed by $\pm 1$ SD of their fitted value. 
\label{tab:ljets_syst_topo_nuisance} } 
\setlength{\tabcolsep}{9pt}
{
\renewcommand{\arraystretch}{1.1} 
\begin{tabular}{c c  c  c  c} \hline \hline
 Source  &   $\sigma_{t\bar{t}}$ [pb] & Offset [pb]     &      $+\sigma$ [pb]    &    $-\sigma$ [pb] \\[1pt] \hline
                                       Statistical only &  7.00  & &  +0.28 &   -0.28 \\ 
\hline  
                                      Muon identification  & &   $-0.02$  &  +0.05 &   $-0.05$   \\ 
                                  Electron identification  & &  +0.14  &  +0.13 &   $-0.12$  \\ 
                                                 Triggers  & &   $-0.08$  &  +0.10 &   $-0.09$  \\ 
                                  Background normalization & &   +0.07   &  +0.06 &   $-0.06$  \\ 
                                          Signal modeling  & &   $-0.22$  &  +0.20 &   $-0.18$  \\ 
                                    Monte Carlo statistics & &   +0.00  &  +0.02 &   $-0.02$   \\ 
                                           MJ background &  &    +0.01 &  +0.00 &   $-0.05$ \\ 
                                                    $f_H$  & &   +0.13  &  +0.03 &   $-0.03$  \\ 
                                          Jet energy scale  & &  +0.26  &  +0.00 &   +0.00  \\ 
                  Jet reconstruction and identification & &    +0.55  &  +0.18 &   $-0.16$  \\ 
                                              Luminosity  & &  +0.45  &  +0.50 &   $-0.44$   \\ 
                                     Template statistics  & &  +0.00   &    +0.04 &   $-0.04$  \\ 
                                                   Other  & &  $-0.01$   &   +0.13 &   $-0.12$  \\ \hline
                                      Total systematics   & &    &    +0.61 &   $-0.55$   \\ 
\hline  
                                             Fit result &  7.68  &  &   +0.71 &   $-0.64$  \\ 
\hline \hline 
 \end{tabular} 
 }
\end{minipage}
\end{center} 
\end{table*} 

The consistency of results between the \ejets\ and \mujets\ channels 
is studied using an ensemble of 10,000 generated pseudo-experiments,  
each representing a single simulation of the results from the data sample,  
assuming $\sigma_{t\bar{t}}$ measured in the combined \ljets\ channel. We vary  the number of signal and background events in each 
pseudo-experiment
within Poisson statistics about their mean values. 
For each pseudo-experiment, we measure  the cross section 
in the \ejets\ and \mujets\ channels by performing a likelihood fit 
in which the parameters corresponding to 
individual sources of systematic uncertainty are varied randomly 
according to Gaussian functions, taking into account 
the correlations between the \ejets\ and \mujets\ channels.  
We record the difference between \sigmatt in both channels and 
calculate, as a measure of consistency, the probability that  
it is equal to or larger than the measured difference   
as shown in  
Table~\ref{tab:ljets_separate_channels_topo_nuisance}. 
The two measurements are found to be 
consistent with a probability of 22\%.

\subsection{$b$-tagging method}

Table~\ref{tab:ljets_separate_channels_btag_nuisance} gives the results of the $b$-tagging method for the \eplus, \muplus, and combined
\lplus\ channels, and 
Table~\ref{tab:ljets_separate_channels_btag_nuisance_err} gives the
systematic uncertainties.  
The consistency of these results is checked with pseudo-experiments 
performed in the same way as described in the previous section. We find that the  \sigmatt  values measured in the  \ejets\ and \mujets\ channels are consistent with a probability of  8\%.

\begin{table}[h]
\newcommand\T{\rule{0pt}{2.6ex}}
\newcommand\B{\rule[-1.2ex]{0pt}{0pt}}
\begin{center}
\caption{Measured $t\bar{t}$ cross section using 
$b$-tagging  for separate and combined \lplus\ channels. 
The first quoted uncertainty denotes the statistical, the second the systematic contribution. The statistical uncertainty is scaled
from the statistical only $\sigma_{t\bar{t}}$ result in Table~\ref{tab:ljets_separate_channels_btag_nuisance_err} to the final $\sigma_{t\bar{t}}$.
The total uncertainty corresponds to the one in the row ``Fit result'' in Table~\ref{tab:ljets_separate_channels_btag_nuisance_err}.
\label{tab:ljets_separate_channels_btag_nuisance} }
\begin{tabular}{lccc} \hline \hline
Channel & $e$+jets & $\mu$+jets & $\ell$+jets \\[1pt]\hline \\[-7pt]
$\sigma_{t\bar{t}}$[pb] & $7.40\pm0.32^{+0.98}_{-0.84}$ & $8.78\pm0.40^{+1.08}_{-0.92}$ & $ 8.13\pm0.25^{+0.99}_{-0.86}$\\[4pt]
\hline\hline
\end{tabular}
\end{center}
\end{table}

\begin{table*}[ht]
\begin{center} 
\begin{minipage}{5.5 in}
 \caption{Measured $t\bar{t}$ cross section and the breakdown of uncertainties 
 for the $b$-tagging method in the \lplus\ channel.  
 The offsets show how the mean value of the measured
 cross section is shifted due to
 each source of systematic uncertainty.  In each line, all but the considered source of systematic
   uncertainty are ignored. 
The $\pm \sigma$ give the impact on the measured
 cross section when the nuisance parameters describing the considered category are 
 changed by $\pm 1$ SD of their fitted value. 
\label{tab:ljets_separate_channels_btag_nuisance_err} }
\setlength{\tabcolsep}{9pt}
{
\renewcommand{\arraystretch}{1.1} 
\begin{tabular}{ccccc} \hline \hline
 Source  &   $\sigma_{t\bar{t}}$ [pb] &  Offset [pb]    &      $+\sigma$ [pb]    &    $-\sigma$ [pb] \\ [1pt] \hline
                                       Statistical only &  7.81  & &  +0.24 &   $-0.24$ \\ \hline
                                    Muon identification  & &  $-0.05$  &  +0.06 &   $-0.05$  \\ 
                                Electron identification  & &  +0.17  &  +0.13 &   $-0.13$  \\ 
                                              Triggers  & &  $-0.13$  &  +0.11 &   $-0.11$  \\ 
                              Background normalization  & &  $-0.00$  &  +0.08 &   $-0.08$  \\ 
                                       Signal modeling  & &  +0.04  &  +0.24 &   $-0.27$  \\ 
				         $b$-tagging & & +0.05  &  +0.34 &   $-0.32$  \\
                               Monte Carlo statistics  & &  $-0.01$  &  +0.09 &   $-0.10$  \\ 
                                         MJ background  & &  $-0.00$  &  +0.06 &   $-0.06$  \\ 
                                                   $f_H$  & &  $-0.04$  &  +0.18 &   $-0.19$  \\ 
                                          Jet energy scale  & &  +0.05  &  +0.09 &   $-0.09$  \\ 
                  Jet reconstruction and identification  & &  +0.02  &  +0.17 &   $-0.16$  \\ 
                                             Luminosity  & &  $-0.02$  &  +0.53 &   $-0.46$  \\ 
                                                  Other  & &  $-0.00$  &  +0.14 &   $-0.13$  \\ 
\hline  
                                      Total systematics  & &    &  +0.77 &   $-0.72$   \\ 
\hline  
                                          Fit result  &  8.13   & &  +1.02 &   $-0.90$  \\ 
\hline \hline 
 \end{tabular} 
 }
\end{minipage}
 \end{center} 
\end{table*}

\subsection{Combined method} 

Table~\ref{tab:ttxsec_ljets_separate_channels_p17p20_0tag1tagtopo_nuisance}
shows results for \sigmatt and $f_H$ in \eplus, \muplus\ and 
\lplus\ channels for the combined method and 
Table~\ref{tab:ttxsec_ljets_systematics_p17p20_0tag1tagtopo_nuisance}
 gives the systematic uncertainties.  
The relative uncertainties on \sigmatt for 
the combined and the kinematic methods are
comparable. This is expected because 
the measurements are
systematically limited. Compared to the kinematic method, the
combined method has improved statistical sensitivity. On the other
hand, we include more sources of systematic uncertainty, such as 
the relatively large $b$-tagging uncertainty, which  reduces slightly  the
final precision.

\begin{table}[h]
\newcommand\T{\rule{0pt}{2.6ex}}
\newcommand\B{\rule[-1.2ex]{0pt}{0pt}}
\begin{center}
\caption{Measured $t\bar{t}$ cross section and the $W$+jets heavy flavor scale factor 
$f_H$  for separate and combined \lplus\ channels, using 
both kinematic information and $b$-tagging.   
The first quoted uncertainty denotes the statistical, the second the systematic contribution. The statistical uncertainty is scaled
from the statistical only $\sigma_{t\bar{t}}$ result in Table~\ref{tab:ttxsec_ljets_systematics_p17p20_0tag1tagtopo_nuisance} to the final $\sigma_{t\bar{t}}$.
The total uncertainty corresponds to the one in the row ``Fit result'' in Table~\ref{tab:ttxsec_ljets_systematics_p17p20_0tag1tagtopo_nuisance}.
\label{tab:ttxsec_ljets_separate_channels_p17p20_0tag1tagtopo_nuisance} }
\begin{tabular}{lccc}\hline\hline
Channel & $e$+jets & $\mu$+jets & $\ell$+jets \\[1pt]\hline \\ [-7pt]
$\sigma_{t\bar{t}}$[pb] & $7.22\pm0.32^{+0.70}_{-0.63}$ & $8.43\pm0.39^{+0.80}_{-0.70}$ & $ 7.78\pm0.25^{+0.73}_{-0.59}$\\ [4pt]
$f_H$      & $1.74\pm0.13^{+0.21}_{-0.21}$ & $1.26\pm0.12^{+0.18}_{-0.17}$ & $ 1.55\pm0.09^{+0.17}_{-0.19}$\\[4pt]
\hline\hline
\end{tabular}
\end{center}
\end{table}

\begin{figure}[h]
\flushleft
\includegraphics[width=0.5\textwidth]{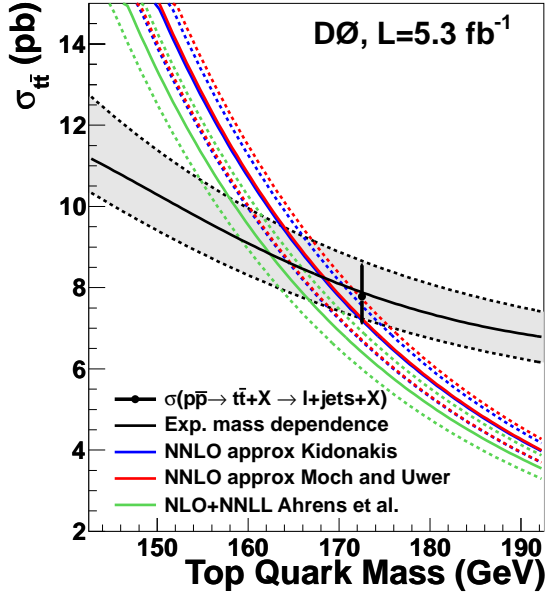}
\begin{center}
\caption{(Color online) Experimental and
theoretical~\cite{SMtheory_A,SMtheory_M,SMtheory_K2} values of \sigmatt\ as a 
function of $m_{t}$.
The point shows  \sigmatt\ measured using the combined method,
the black line the fit with Eq.~\ref{eq_massfit}, and the gray band with
its dashed delimiting lines the corresponding total experimental uncertainty.
Each  curve is bracketed by dashed lines of the
corresponding color that represent the theoretical uncertainties
due to the choice of PDF and the renormalization
and factorization scales (added linearly).
\label{fig:mass}}
\end{center}
\end{figure}

\begin{table*}[ht]
\begin{center} 
\begin{minipage}{5.5 in}
 \caption{Measured $t\bar{t}$ cross section and the breakdown of uncertainties 
 for the combined kinematic and $b$-tagging method in the \lplus\ channel. 
 The offsets show how the mean value of the measured
 cross section is shifted due to
 each source of systematic uncertainty. In each line, all but the considered source of systematic
   uncertainty are ignored. 
The $\pm \sigma$ give the impact on the measured
 cross section when the nuisance parameters describing the considered category are 
 changed by $\pm 1$ SD of their fitted value. 
\label{tab:ttxsec_ljets_systematics_p17p20_0tag1tagtopo_nuisance} } 
\setlength{\tabcolsep}{9pt}
{
\renewcommand{\arraystretch}{1.1} 
\begin{tabular}{ccccc} \hline\hline
 Source  &   $\sigma_{t\bar{t}}$ [pb] &  Offset [pb]   &      $+\sigma$ [pb]    &    $-\sigma$ [pb] \\ [1pt] \hline
                                                Statistical only &  7.58 & &  +0.24 &   $-0.24$ \\ 
\hline  
                                    Muon identification  & &  $-0.04$  &  +0.05 &   $-0.05$  \\ 
                                   Electron identification   & &  +0.14  &  +0.12 &   $-0.12$  \\ 
                                                  Triggers  & &  -$0.09$  &  +0.09 &   $-0.11$  \\ 
                                  Background normalization  & &  +0.00   &   +0.07 &   $-0.06$  \\ 
                                          Signal modeling  & &  $-0.06$  &  +0.23 &   $-0.21$  \\ 
                                               $b$-tagging  & &  $-0.14$  &  +0.12 &   $-0.12$  \\ 
                                 Monte Carlo statistics  & &  $-0.01$  &  +0.06 &   $-0.06$  \\ 
                                     Fake background  & &  $-0.01$  &  +0.06 &   $-0.04$  \\ 
                                           $f_H$  & &  $-0.00$  &  +0.02 &   $-0.02$  \\ 
                                           Jet energy scale  & &  $-0.03$  &  +0.00 &   $-0.00$  \\ 
                  Jet reconstruction and identification  & &  +0.18  &  +0.18 &   $-0.17$  \\ 
                                              Luminosity  & &  +0.12  &  +0.51 &   $-0.44$  \\ 
                                     Template statistics  & & +0.00 & +0.03 & $-0.03$ \\ 
                                                   Other  & &  +0.01  &  +0.14 &   $-0.13$  \\ 
\hline  
                                       Total systematics  & &    &  +0.65 &   $-0.58$   \\ 
\hline  
                                               Fit result  &  7.78   & &  +0.77 &   $-0.64$  \\ 
\hline \hline 
 \end{tabular} 
}
\end{minipage}
 \end{center} 
\end{table*}

\subsection{Top quark mass dependency for the combined method} \label{sec:massdep} 

Different selection efficiencies lead to a dependence of \sigmatt 
on $m_t$. This is studied using 
simulated samples of \ttbar\ events generated at different values of
$m_t$ using the \alpgen\ event generator followed by \pythia\, for the simulation of parton-shower development. The resulting measurements 
are summarized in Table~\ref{tab:xsvsmass} 
and can be parametrized as a function of $m_{t}$ as  
\begin{eqnarray} \label{eq_massfit}
\lefteqn{\sigma_{t\bar{t}}(m_t) =}  \\   \nonumber
& &  \frac{1}{m^4_{t}} \left[ a + b (m_{t} -m_0) + c(m_{t} -m_0)^2 + d  (m_{t} -m_0)^3 \right ],
\end{eqnarray}
where $\sigma_{t\bar{t}}$ and $m_{t}$ are in~pb and~GeV, respectively, and 
$m_0 = 170$~GeV, $a = 5.78874  \times 10^{9}{\rm ~pb~GeV}^{4}, b = -4.50763 \times 10^{7}{\rm ~pb~GeV}^{3}, c =
1.50344 \times 10^{5}{\rm ~pb~GeV}^{2}$, and $d = - 1.00182 \times 10^{3}{\rm ~pb~GeV}$.

\begin{table}[h]
\newcommand\T{\rule{0pt}{2.6ex}}
\newcommand\B{\rule[-1.2ex]{0pt}{0pt}}
\begin{center}
\caption{The  $t\bar{t}$ cross sections measured using the combined method for different assumed 
top quark masses. The uncertainty is the combined statistical plus systematic uncertainty.   
\label{tab:xsvsmass} }
\begin{tabular}{lc} \hline \hline
$m_{t}$ & $\sigma_{t\bar{t}}$[pb] \\[1pt]\hline
\\[-7pt]
150   & $10.27^{+1.10}_{-0.88} $ \\[2pt]
160   & $9.14^{+0.86}_{-0.79} $\\ [2pt]
165   & $8.56^{+0.82}_{-0.71} $\\ [2pt]
170   & $8.09^{+0.77}_{-0.68} $\\ [2pt]
172.5 & $7.78^{+0.77}_{-0.64} $\\ [2pt]
175   & $7.65^{+0.79}_{-0.62} $\\ [2pt]
180   & $7.46^{+0.74}_{-0.61} $\\ [2pt]
185   & $7.06^{+0.67}_{-0.60} $\\ [2pt]
190   & $6.85^{+0.66}_{-0.62} $\\ [2pt]\hline\hline
\end{tabular}
\end{center}
\end{table}

In Fig.~\ref{fig:mass} we compare this parameterization to three  approximations 
to \sigmatt at 
next-to-next-to-leading-order (NNLO) QCD that include
all next-to-next-to-leading logarithms (NNLL) in NNLO
QCD~\cite{SMtheory_A,SMtheory_M,SMtheory_K2}. 


\section{Conclusion}
We  measured the \ttbar\ production cross section in the 
$\ell$+jets final states using different analysis techniques. In
\lumi\ of integrated luminosity collected with the D0 detector,  
for a top quark mass of 172.5 GeV, we obtain:
\begin{equation}
\sigma_{t\overline{t}} = 
\combiresult \combitotal \:{\rm (stat+syst+lumi)}\:{\rm pb}, \nonumber
\end{equation}
using both kinematic event information and $b$-jet identification and 
simultaneously measuring the cross section and 
the ratio of $W$+heavy flavor jets to $W$+light flavor jets. 
The precision achieved is approximately
$9$\%. A result of similar precision from  the 
CDF Collaboration is available in Ref.~\cite{cdf_xsec}. 
All our results are consistent with the 
theoretical predictions of 
$\sigma_{t\bar{t}}=6.41^{+0.51}_{-0.42}$\,pb~\cite{SMtheory_A} and  
$\sigma_{t\bar{t}}=7.46^{+0.48}_{-0.67}$\,pb~\cite{SMtheory_M}.

\section*{Acknowledgments}
%
We thank the staffs at Fermilab and collaborating institutions,
and acknowledge support from the
DOE and NSF (USA);
CEA and CNRS/IN2P3 (France);
FASI, Rosatom and RFBR (Russia);
CNPq, FAPERJ, FAPESP and FUNDUNESP (Brazil);
DAE and DST (India);
Colciencias (Colombia);
CONACyT (Mexico);
KRF and KOSEF (Korea);
CONICET and UBACyT (Argentina);
FOM (The Netherlands);
STFC and the Royal Society (United Kingdom);
MSMT and GACR (Czech Republic);
CRC Program and NSERC (Canada);
BMBF and DFG (Germany);
SFI (Ireland);
The Swedish Research Council (Sweden);
and
CAS and CNSF (China).
%


\newpage


\begin{thebibliography}{99}
\bibitem{eta} The rapidity $y$ and pseudorapidity $\eta$ are defined as functions
    of the polar angle $\theta$ and parameter $\beta$ as
    $y(\theta,\beta) \equiv
    {\frac{1}{2}} \ln{[(1+\beta\cos{\theta})/(1-\beta\cos{\theta})]}$ and
    $\eta(\theta) \equiv y(\theta,1)$, where $\beta$ is the ratio of a particle's 
    momentum to its energy. We distinguish detector $\eta$ ($\eta_{det}$) and physics
    $\eta$, where the former is defined with respect to the center of the
    detector and the latter with respect to the $p\bar{p}$ interaction vertex.
\bibitem{ip} The impact parameter is defined as the distance of closest 
    approach ($d_{ca}$) of the track to the PV in the 
    plane transverse to the beamline. The impact parameter significance is defined as
    $d_{ca}/\sigma_{d_{ca}}$, where $\sigma_{d_{ca}}$ is the uncertainty 
    on $d_{ca}$.
\end{thebibliography}
\end{document}